\documentclass[pdflatex,sn-mathphys-num]{sn-jnl}

\usepackage{breqn}
\usepackage{hyperref}
\usepackage{graphicx}%
\usepackage{multirow}%
\usepackage{amsmath,amssymb,amsfonts}%
\usepackage[numbers]{natbib}%
\usepackage{amsthm}%
\usepackage{mathrsfs}%
\usepackage[title]{appendix}%
\usepackage{xcolor}%
\usepackage{textcomp}%
\usepackage{manyfoot}%
\usepackage{booktabs}%
\usepackage{algorithm}%
\usepackage{algorithmicx}%
\usepackage{algpseudocode}%
\usepackage{listings}%

\begin{document}

\title[Article Title]{Manifestation Of Quantum Forces In Spacetime: Towards A General Theory Of Quantum Forces}

\author[1]{\fnm{Raheem} \sur{Adom}}
\affil[1]{ \orgaddress{\city{Accra}, \country{Ghana}}}


\abstract{This study introduces the quantum force wave equation (QFWE) as a general theory of quantum forces, a novel framework that redefines quantum forces as emergent phenomena arising from the interaction between quantum particles and curved spacetime. By coupling wavefunctions to spacetime curvature and gauge fields, the theory establishes a dynamic, bidirectional relationship between quantum states and spacetime geometry. This approach provides a unified description of quantum forces in highly curved and dynamic gravitational fields, extending beyond the limitations of existing theories. The theory offers fresh insights into quantum gravity, quantum field theory in curved spacetime, and particle physics in extreme conditions, serving as a versatile tool for exploring the interplay between quantum mechanics and spacetime structure. This work lays the foundation for the advancement of high-energy physics and cosmology in regimes where spacetime curvature is fundamental.}




\maketitle

\section{Introduction}\label{1}
The study of forces emerging from the quantum behavior of particles remains a significant frontier in theoretical physics, particularly in the context of their interaction with curved spacetime. Although classical mechanics defines force through Newton’s laws and quantum mechanics reinterprets it via interaction potentials and gauge fields, the introduction of spacetime curvature fundamentally alters this picture. In flat spacetime, quantum forces are effectively described by quantum field theory (QFT). However, in curved spacetime—such as near black holes, neutron stars, or during the early universe—new complexities arise as the interplay between quantum states and spacetime geometry introduces effects that neither standard quantum mechanics nor general relativity fully account for. This raises foundational questions about how quantum forces manifest in such environments.

Reconciling quantum mechanics with spacetime curvature remains a central challenge in modern physics. Historically, general relativity and quantum mechanics have been regarded conceptually incompatible, with attempts to unify them often leading to inconsistencies \cite{1}. Quantum field theory in curved spacetime provides a semi-classical framework for studying quantum fields in fixed curved geometries but neglects the dynamic interaction between quantum particles and spacetime itself. Quantum gravity theories, such as string theory and loop quantum gravity, aim to unify quantum mechanics and general relativity by quantizing spacetime, yet they remain abstract and computationally complex. Despite these advances, a comprehensive description of quantum forces in curved spacetime, especially at the particle level, remains elusive, leaving gaps in understanding scenarios where both quantum and gravitational effects are significant.

To address these challenges, this study introduces the quantum force wave equation (QFWE), which redefines quantum forces as emergent phenomena arising from the interaction of wavefunctions with spacetime curvature and gauge fields \cite{2}. By embedding wavefunctions within a dynamically curved spacetime framework, the QFWE incorporates covariant derivatives to account for gravitational effects and gauge terms to describe field interactions. This unified approach enables a consistent description of quantum forces across a wide range of spacetime geometries, from flat regions to highly curved environments.

The QFWE represents a critical advancement toward bridging quantum mechanics and general relativity by redefining force as a property that emerges from the dynamic coupling of quantum states to spacetime geometry. This framework extends the understanding of quantum forces, offering a versatile tool for studying their role in high-energy physics, cosmology, and particle dynamics in extreme conditions. It provides a theoretical foundation for exploring the intricate interplay between quantum fields and spacetime, with implications for quantum gravity, black hole physics, and beyond.

Importantly, this framework provides a solution to the so-called backreaction effect which entails the influence of quantum fields to the curvature of spacetime itself. This is achieved by embedding the QFWE to the Einstein's field equations (EFE). This approach modifies the Einstein's tensor which introduces a novel way to think of gravity at the quantum scale. Unlike general relativity which explain the geometry of spacetime as a classical backdrop influenced by the presence of matter and energy, thus showing how the geometry of spacetime dictates the motion of massive bodies, the modified EFE shows how the quantum nature of spacetime geometry dictates the motion of quantum fields. 

The remainder of this paper is organized as follows. In Section ~\ref{2}, we outline the theoretical framework that underpins the QFWE, describing the key mathematical structures and physical concepts involved. Section ~\ref{3} provides a detailed derivation of the equation. Section ~\ref{4} discusses the physical interpretation of the QFWE, offering insights into its implications for quantum dynamics in curved spacetime. Section ~\ref{5} provides potential applications of the established frame work in black hole physics, quantum information, quantum materials, and high-energy physics. In Section ~\ref{6}, we explore potential observational signatures and experimental feasibility, focusing on astrophysical, cosmological, and quantum contexts. Finally, in Section ~\ref{7}, we summarize our findings and propose directions for future research aimed at developing a unified theory of quantum forces.

This work seeks to establish a foundation for addressing open questions in quantum mechanics and general relativity, advancing our understanding of quantum forces as a fundamental aspect of the interaction between matter and spacetime.

\section{Theoretical Framework}\label{2}
To investigate the behavior of quantum forces in a curved spacetime, we constructed a framework that integrates quantum mechanics, gauge theory, and general relativity. The central element of this framework is the QFWE, which describes the influence of spacetime curvature and gauge fields on the evolution of a particle’s wave function. This equation provides a generalized form of the quantum force, making it applicable to a wide range of quantum systems embedded in various spacetime geometries.\\
The equation is structured around a generalized force term \(F_{\mu}\) represented as
\begin{equation}
F_{\mu} = \left( F_{0},\mathbf{F}\right).\label{1}
\end{equation}
Where \(F_{0}\) represents the generalized time component and \(\textbf{F}\in {\mathbb {R}}^3\) represents the spatial force components. In this study, we introduce a specific form of \(F_{0}\) which makes the time component energy-dependent as 
\begin{equation}
F_{0}=\frac{E^2}{hc}.\label{2}
\end{equation}
Where \(E\) is the relativistic energy of the quantum particles propagating through spacetime, \(h\) is Planck's constant and \(c\) is the speed of light.  This quadratic energy dependence and presence of the Planck's constant enables \(F_{\mu}\) to describe interactions beyond classical force definitions.
The wave function, describing quantum particles of various spins, is defined in this framework as:
\begin{equation}
\Psi_{\alpha_1, \dots, \alpha_n} = \mathcal{P} \exp \left( i \int_{\mathcal{C}} c_{\mu} \, dx^{\mu} \right) \chi_{\alpha_1, \dots, \alpha_n}.\label{3}
\end{equation}
In the above equation, \(\mathcal{P}\) is the path-ordering operator that ensures that the gauge fields and gravitational connections are properly ordered along paths \(\mathcal{C}\) \cite{3} and \(\chi_{\alpha_1 , \dots, \alpha_n}\) is the tensor-spinor field that describes the intrinsic spin of quantum particles and can accommodate various spin representations \cite{4,5}. The term \(c_{\mu}\) within the path-ordered exponential is crucial as it encapsulates the interactions between the wavefunction and all fundamental fields. It serves as the \textit{field-coupling vector}, which is taken to be of the form
\begin{equation}
c_{\mu}=k_{\mu} + g_a \, A_{\mu}^a T^a+\omega_{\mu}^{ab} J_{ab},\label{4}
\end{equation}
where \(k_{\mu}\) represents the wave vector which describes the free propagation of particles in spacetime, \(g_a \, A_{\mu}^a T^a\) represents the gauge field term that accounts for the interaction with both the Abelian and non-Abelian gauge fields, \(\omega_{\mu}^{ab}\) represents the spin connection in curved spacetime and \(J_{ab}\) represent the generators of the Lorentz group \(SO(1,3)\) \cite{6,7} (or a generalization depending on the spacetime symmetry group) acting on the spinor and tensor indices. 

To ensure dimensional consistency across all terms, we note that \(g_a\), \(T_a\), and \(J_{ab}\) are all dimensionless, while \(A^a_{\mu}\) and \(\omega^{ab}_{\mu}\) share the same dimensionality as \(k_{\mu}\). Thus, \(c_{\mu}\) has a consistent dimension of \([L^{-1}]\), aligning with its role as a covariant vector that governs particle interactions in curved spacetime.

This construction guarantees that \(c_{\mu}\) is mathematically well-defined and physically meaningful, enabling the unification of quantum field interactions and spacetime curvature in the proposed framework.

The expression \(\mathcal{P} \exp \left( i \int_{\mathcal{C}} c_{\mu} dx^{\mu}\right)\) serves as a path-ordered exponential in the wave function which represents the total phase accumulated over path \(\mathcal{C}\) with \(dx^{\mu}\) as is the infinitesimal distance along the path. We denote this as;
\begin{equation}
\psi=\mathcal{P} \exp \left( i \int_{\mathcal{C}} c_{\mu} dx^{\mu}\right).\label{5}
\end{equation}
In this study, we consider the covariant derivative \(\nabla_{\mu}\) of the path-ordered exponential \(\psi\) as
\begin{equation}
\nabla_{\mu}\psi=ic_{\mu}\psi+z_{\mu}\psi,\label{6}
\end{equation}
where \(z_{\mu}\) is a general term that accounts for all complexities beyond the leading \(ic_{\mu}\) contribution. This coupling vector \(z_{\mu}\) is inherently non-local because it depends on integrals over the entire path \(\mathcal{C}\). The non-local coupling vector \(z_{\mu}\) in its general representation which is an infinite series of higher order corrections, where each term includes nested commutators integrated over the path, is considered in this study to take the form
\begin{equation}
z_\mu = \sum_{n=1}^\infty \int_{\mathcal{C}} \cdots \int_{\mathcal{C}} \mathcal{P} \Big[ c_\mu(x), \big[c_{\nu_1}(x_1), \big[c_{\nu_2}(x_2), \dots, c_{\nu_n}(x_n)\big]\big]\Big] dx_1^{\nu_1}, dx_2^{\nu_2}\cdots dx_n^{\nu_n}.\label{7}
\end{equation}
where \(\Big[ c_\mu(x), \big[c_{\nu_1}(x_1), \big[c_{\nu_2}(x_2), \dots, c_{\nu_n}(x_n)\big]\big]\Big]\) contain commutators at increasing nesting levels. Each commutator reflects the non-Abelian nature of the fields encoded in the field coupling vector \(c_{\mu}(x)\) where the non-Abelian nature of gravitational effects generally arises directly from \(SU(N)\) gauge group (coupled to conformal gravity)  analogous to how non-Abelian gauge fields behave in Yang-Mills theory \cite{8,9}. The commutators are integrated over all points \(x_1^{\nu_1},x_2^{\nu_2}\cdots x_n^{\nu_n}\). This emphasizes the non-locality of \(z_{\mu}\), as its value depends on the entire path, not just the local behavior of \(c_{\mu}\) at a single point.\\ 
Substituting Eq. \ref{6} into Eq. \ref{3}, yields the following:
\begin{equation}
\Psi_{\alpha_1 , \dots, \alpha_n} = \psi \chi_{\alpha_1 , \dots, \alpha_n}.\label{8}
\end{equation}
This is a well-motivated generalization of quantum particles propagating in all the known force fields. The equivalence of Euclidean and Lorentzian formulations has been grounded in standard theories, and it is not clear how such a correspondence would be possible for higher derivative theories \cite{10} especially this one. However, further study of the theory could offer a way to show significant correspondence because it is a Lorentz covariant. 

The resulting equation relates \(F_{\mu}\) and the wave function through an operator \(R_{\mu}\). The operator \(R_{\mu}\), referred to in this study as the General Quantum Manipulator (GQM) encapsulates the combined effects of spacetime curvature, gauge field interactions, and non-local quantum corrections on the evolution of a quantum particle’s wavefunction. Mathematically, \(R_{\mu}\) is constructed using covariant derivatives that account for both the intrinsic spin of the particle and the external fields present in the spacetime geometry. It integrates higher-order derivatives of the wavefunction and the tensor-spinor field \(\chi_{\alpha_1, \dots, \alpha_n}\), capturing the coupling between quantum states and spacetime dynamics.

The structure of \(R_{\mu}\) is designed to unify local and non-local contributions to quantum forces. Local effects arise through the coupling of the wavefunction to gauge fields and spin connections, while non-locality is introduced via the \(z_{\mu}\) term, which represents nested commutators along spacetime paths. This term reflects the non-Abelian nature of gauge and gravitational fields, allowing \(R_{\mu}\) to describe quantum correlations and long-range interactions.

In essence, \(R_{\mu}\) serves as a comprehensive operator that governs the dynamics of quantum particles under the influence of spacetime curvature, gauge field interactions, and quantum spin effects. Its inclusion in the QFWE ensures that the framework remains consistent with general relativity, gauge invariance, and quantum mechanical principles, making it a versatile tool for describing quantum forces in curved spacetime.

\section{Mathematical Analysis}\label{3}
\subsection{Derivation Of The QFWE}
From Newton's second law of motion, the force is equal to the time rate of change of momentum, which is given by:
\begin{equation}
F = \frac{dp}{dt}.\label{9}
\end{equation}
Where \(p\) is the momentum and \(t\) is the time.\\
The energy wave equation for a matter wave is given as \(E = \hbar \omega\). Where \(\hbar\) is the rationalized Planck's constant and \(\omega\) is the angular frequency of the matter wave. But from De Broglie's wave-particle equation, \(p= \frac{h}{\lambda}\) which can also be written as \(p= \hbar k\) where \(k\) is the wave number. Solving for \(\hbar\) gives \(\hbar = \frac{p}{k}\). Substituting this into the wave equation yields: \(E=\frac{p \omega}{k}\). Solving for \(p\) yields \(p = \frac{E k}{\omega}\). Substituting this equation into Eq. \ref{10} yields:
\begin{equation}
F=\frac{d(\frac{E k}{\omega})}{dt}.\label{10}
\end{equation}
In this study, we consider the mechanical energy \(E\) of quantum particles to vary across spaces to obtain the force. The mechanical energy is taken to be time-independent and is given by
\begin{equation}
E=\frac{\hbar^{2}}{2m}k^{2}+V.\label{11}
\end{equation}
Where \(\frac{\hbar^{2}}{2m}k^{2}\) is the kinetic energy and \(V\) is the potential energy like in the Schrödinger equations. This is equivalent to the matter wave equation introduce before as \(E=\hbar \omega\) which is time-dependent because of the presence of the angular frequency \(\omega\). We can see that the mechanical energy equation for a matter wave can be both time dependent and independent. This is a way of avoiding the complexities that may arise when using the product rule when the total energy is considered to be time-dependent. The angular velocity is given by \(\omega=\frac{2 \pi}{t}\). Substituting it into Eq. \ref{10} results in
\begin{equation}
F=\frac{d(\frac{E k}{2 \pi}t)}{dt}.\label{12}
\end{equation}
Solving the differential equation results to;
\begin{equation}
F=\frac{E k}{2 \pi}.\label{13}
\end{equation}
But \(E=\hbar \omega\);
\begin{equation}
F=\frac{\hbar \omega k}{2 \pi}.\label{14}
\end{equation}
Multiplying the left-hand side of Eq. \ref{14} by \(\frac{k}{k}\);
\begin{equation}
F=\frac{\hbar \omega k^{2}}{2 \pi k}.\label{15}
\end{equation}
However, \(\frac{\omega}{k} = v\), where \(v\) is velocity. With this, the Eq. \ref{15} becomes;
\begin{equation}
F=\frac{\hbar v k^{2}}{2 \pi}.\label{16}
\end{equation}
But from the momentum equation \(p=mv\), \(v=\frac{p}{m}\). By substituting this into Eq. \ref{16} results in
\begin{equation}
F= \frac{\hbar p k^{2}}{2 \pi m}.\label{17}
\end{equation}
From the De Broglie wave-particle equation as given earlier, \(p=\hbar k\). By substituting this into Eq. \ref{17} results in
\begin{equation}
F=\frac{\hbar^{2}}{2 \pi m} k^{3}.\label{18}
\end{equation}
The Eq. \ref{18} yields the \textit{quantum mechanical version} of Newton's second law of motion: However we must use it in a covariant formulation. Considering this, we can consider its covariant form as:
\begin{equation}
F_{\mu}=\frac{\hbar^{2}}{2 \pi m}k_{\nu}k^{\nu}k_{\mu}.\label{19}
\end{equation} 
Where \(F_{\mu}\) is the force vector term and \(k_{\nu}k^{\nu}k_{\mu}\) are the wave vectors. However, to account for all field interactions, we need to extend the equation to incorporate gauge and gravitational field effects by replacing the wave vector with the field-coupling vector. With this, we can write its extension as
\begin{equation}
F_{\mu}=\frac{\hbar^{2}}{2 \pi m}c_{\nu}c^{\nu}c_{\mu}.\label{20}
\end{equation}
In this study, we would choose \(\nabla_{\mu}\) to represent the covariant derivative on any manifold with any type of connection whether it is a Levi-Civita, a general affine connection, or a connection in principal or vector bundle. The first derivative of the wavefunction yields
\begin{equation}
\nabla_{\mu}\Psi_{\alpha_1 , \dots, \alpha_n}=\nabla_{\mu}(\psi\chi_{\alpha_1 , \dots, \alpha_n}).\label{21}
\end{equation}
Applying the product rule to Eq. \ref{21} yields
\begin{equation}
\nabla_{\mu}\Psi_{\alpha_1 , \dots, \alpha_n}=\chi_{\alpha_1 , \dots, \alpha_n}\nabla_{\mu}\psi+\psi\nabla_{\mu}\chi_{\alpha_1 , \dots, \alpha_n}.\label{22}
\end{equation}
With the first derivative obtained, we now move to the second derivative with, as follows:
\begin{equation}
\nabla^{\nu}\nabla_{\mu}\Psi_{\alpha_1 , \dots, \alpha_n}=\nabla^{\nu}(\chi_{\alpha_1 , \dots, \alpha_n}\nabla_{\mu}\psi+\psi\nabla_{\mu}\chi_{\alpha_1 , \dots, \alpha_n}).\label{23}
\end{equation}
\begin{equation}
\nabla^{\nu}\nabla_{\mu}\Psi_{\alpha_1 , \dots, \alpha_n}=\nabla^{\nu}(\chi_{\alpha_1 , \dots, \alpha_n}\nabla_{\mu}\psi)+\nabla^{\nu}(\psi\nabla_{\mu}\chi_{\alpha_1 , \dots, \alpha_n}).\label{24}
\end{equation}
Applying the product rule to Eq. \ref{24} yields;
\begin{dmath}
\nabla^{\nu}\nabla_{\mu}\Psi_{\alpha_1 , \dots, \alpha_n}=\nabla^{\nu}\chi_{\alpha_1 , \dots, \alpha_n}\nabla_{\mu}\psi+(\nabla^{\nu}\nabla_{\mu}\psi)\chi_{\alpha_1 , \dots, \alpha_n}+\nabla^{\nu}\psi\nabla_{\mu}\chi_{\alpha_1 , \dots, \alpha_n}+
\\
(\nabla^{\nu}\nabla_{\mu}\chi_{\alpha_1 , \dots, \alpha_n})\psi.\label{25}
\end{dmath}
With the second derivative obtained, we can now move to the third derivative, given as:
\begin{dmath}
\nabla_{\nu}\nabla^{\nu}\nabla_{\mu}\Psi_{\alpha_1 , \dots, \alpha_n}=\nabla_{\nu}(\nabla^{\nu}\chi_{\alpha_1 , \dots, \alpha_n}\nabla_{\mu}\psi+(\nabla^{\nu}\nabla_{\mu}\psi)\chi_{\alpha_1 , \dots, \alpha_n}+\nabla^{\nu}\psi\nabla_{\mu}\chi_{\alpha_1 , \dots, \alpha_n}+
\\
(\nabla^{\nu}\nabla_{\mu}\chi_{\alpha_1 , \dots, \alpha_n})\psi).\label{26}
\end{dmath}
\begin{dmath}
\nabla_{\nu}\nabla^{\nu}\nabla_{\mu}\Psi_{\alpha_1 , \dots, \alpha_n}=\nabla_{\nu}(\nabla^{\nu}\chi_{\alpha_1 , \dots, \alpha_n}\nabla_{\mu}\psi)+\nabla_{\nu}((\nabla^{\nu}\nabla_{\mu}\psi)\chi_{\alpha_1 , \dots, \alpha_n})+\nabla_{\nu}(\nabla^{\nu}\psi\nabla_{\mu}\chi_{\alpha_1 , \dots, \alpha_n}+\nabla_{\nu}((\nabla^{\nu}\nabla_{\mu}\chi_{\alpha_1 , \dots, \alpha_n})\psi)).\label{27}
\end{dmath}
Applying the product rule to Eq. \ref{27} yields
\begin{dmath}
\nabla_{\nu}\nabla^{\nu}\nabla_{\mu}\Psi_{\alpha_1 , \dots, \alpha_n}=(\nabla_{\nu}\nabla^{\nu}\chi_{\alpha_1 , \dots, \alpha_n})\nabla_{\mu}\psi+\nabla^{\nu}\chi_{\alpha_1 , \dots, \alpha_n}(\nabla_{\nu}\nabla_{\mu}\psi)+(\nabla_{\nu}\nabla^{\nu}\nabla_{\mu}\psi)\chi_{\alpha_1 , \dots, \alpha_n}+\nabla_{\nu}\chi_{\alpha_1 , \dots, \alpha_n}(\nabla^{\nu}\nabla_{\mu}\psi)+(\nabla_{\nu}\nabla^{\nu}\psi)\nabla_{\mu}\chi_{\alpha_1 , \dots, \alpha_n}+\nabla^{\nu}\psi(\nabla_{\nu}\nabla_{\mu}\chi_{\alpha_1 , \dots, \alpha_n})+(\nabla_{\nu}\nabla^{\nu}\nabla_{\mu}\chi_{\alpha_1 , \dots, \alpha_n})\psi+\nabla_{\nu}\psi(\nabla^{\nu}\nabla_{\mu}\chi_{\alpha_1 , \dots, \alpha_n}).\label{28}
\end{dmath}
 Now, we must solve for each term the covariant derivative acting on the path-ordered exponential \(\psi\). The solution for each term yields.
\begin{equation}
(\nabla_{\nu}\nabla^{\nu}\chi_{\alpha_1 , \dots, \alpha_n})\nabla_{\mu}\psi=i(\nabla_{\nu}\nabla^{\nu}\chi_{\alpha_1 , \dots, \alpha_n})c_{\mu}\psi+(\nabla_{\nu}\nabla^{\nu}\chi_{\alpha_1 , \dots, \alpha_n})z_{\mu}\psi.\label{29}
\end{equation}
\begin{dmath}
\nabla^{\nu}\chi_{\alpha_1 , \dots, \alpha_n}(\nabla_{\nu}\nabla_{\mu}\psi)=i(\nabla^{\nu}\chi_{\alpha_1 , \dots, \alpha_n}\nabla_{\nu}c_{\mu})\psi-\nabla^{\nu}\chi_{\alpha_1 , \dots, \alpha_n} c_{\nu}c_{\mu}\psi+i(\nabla^{\nu}\chi_{\alpha_1 , \dots, \alpha_n} z_{\nu}c_{\mu})\psi+(\nabla^{\nu}\chi_{\alpha_1 , \dots, \alpha_n}\nabla_{\nu}z_{\mu})\psi+i(\nabla^{\nu}\chi_{\alpha_1 , \dots, \alpha_n} c_{\nu}z_{\mu})\psi+(\nabla^{\nu}\chi_{\alpha_1 , \dots, \alpha_n} z_{\nu}z_{\mu})\psi.\label{30}
\end{dmath}
\begin{dmath}
(\nabla_{\nu}\nabla^{\nu}\nabla_{\mu}\psi)\chi_{\alpha_1 , \dots, \alpha_n}=i(\nabla_{\nu}\nabla^{\nu}c_{\mu})\Psi_{\alpha_1 , \dots, \alpha_n}-c_{\nu}\nabla^{\nu}c_{\mu}\Psi_{\alpha_1 , \dots, \alpha_n}+iz_{\nu}\nabla^{\nu}c_{\mu}\Psi_{\alpha_1 , \dots, \alpha_n}-\nabla^{\nu}(c^{\nu}c_{\mu})\Psi_{\alpha_1 , \dots, \alpha_n}-ic_{\nu}c^{\nu}c_{\mu}\Psi_{\alpha_1 , \dots, \alpha_n}-z_{\nu}c^{\nu}c_{\mu}\Psi_{\alpha_1 , \dots, \alpha_n}+i\nabla_{\nu}(z^{\nu}c_{\mu})\Psi_{\alpha_1 , \dots, \alpha_n}-c_{\nu}z^{\nu}c_{\mu}\Psi_{\alpha_1 , \dots, \alpha_n}+iz_{\nu}z^{\nu}c_{\mu}\Psi_{\alpha_1 , \dots, \alpha_n}+(\nabla_{\nu}\nabla^{\nu}z_{\mu})\Psi_{\alpha_1 , \dots, \alpha_n}+ic_{\nu}\nabla^{\nu}z_{\mu}\Psi_{\alpha_1 , \dots, \alpha_n}+z_{\nu}\nabla^{\nu}z_{\mu}\Psi_{\alpha_1 , \dots, \alpha_n}+i\nabla_{\nu}(c^{\nu}z_{\mu})\Psi_{\alpha_1 , \dots, \alpha_n}-c_{\nu}c^{\nu}z_{\mu}\Psi_{\alpha_1 , \dots, \alpha_n}+iz_{\nu}c^{\nu}z_{\mu}\Psi_{\alpha_1 , \dots, \alpha_n}+\nabla_{\nu}(z^{\nu}z_{\mu})\Psi_{\alpha_1 , \dots, \alpha_n}+ic_{\nu}z^{\nu}z_{\mu}\Psi_{\alpha_1 , \dots, \alpha_n}+z_{\nu}z^{\nu}z_{\mu}\Psi_{\alpha_1 , \dots, \alpha_n}.\label{31}
\end{dmath}
\begin{dmath}
\nabla_{\nu}\chi_{\alpha_1 , \dots, \alpha_n}(\nabla^{\nu}\nabla_{\mu}\psi)=i(\nabla_{\nu}\chi_{\alpha_1 , \dots, \alpha_n}\nabla^{\nu}c_{\mu})\psi-\nabla_{\nu}\chi_{\alpha_1 , \dots, \alpha_n} c^{\nu}c_{\mu}\psi+i(\nabla_{\nu}\chi_{\alpha_1 , \dots, \alpha_n} z^{\nu}c_{\mu})\psi+(\nabla_{\nu}\chi_{\alpha_1 , \dots, \alpha_n}\nabla^{\nu}z_{\mu})\psi+i(\nabla_{\nu}\chi_{\alpha_1 , \dots, \alpha_n} c^{\nu}z_{\mu})\psi+(\nabla_{\nu}\chi_{\alpha_1 , \dots, \alpha_n} z^{\nu}z_{\mu})\psi.\label{32}
\end{dmath}
\begin{dmath}
(\nabla_{\nu}\nabla^{\nu}\psi)\nabla_{\mu}\chi_{\alpha_1 , \dots, \alpha_n}=i(\nabla_{\nu}c^{\nu}\nabla_{\mu}\chi_{\alpha_1 , \dots, \alpha_n})\psi-(c_{\nu}c^{\nu}\nabla_{\mu}\chi_{\alpha_1 , \dots, \alpha_n})\psi+i(z_{\nu}c^{\nu}\nabla_{\mu}\chi_{\alpha_1 , \dots, \alpha_n})\psi+(\nabla_{\nu}z^{\nu}\nabla_{\mu}\chi_{\alpha_1 , \dots, \alpha_n})\psi+i(c_{\nu}z^{\nu}\nabla_{\mu}\chi_{\alpha_1 , \dots, \alpha_n})\psi+(z_{\nu}z^{\nu}\nabla_{\mu}\chi_{\alpha_1 , \dots, \alpha_n})\psi.\label{33}
\end{dmath}
\begin{equation}
\nabla^{\nu}\psi(\nabla_{\nu}\nabla_{\mu}\chi_{\alpha_1 , \dots, \alpha_n})=i(c^{\nu}\nabla_{\nu}\nabla_{\mu}\chi_{\alpha_1 , \dots, \alpha_n})\psi+(z^{\nu}\nabla_{\nu}\nabla_{\mu}\chi_{\alpha_1 , \dots, \alpha_n})\psi.\label{34}
\end{equation}
\begin{equation}
\left(\nabla_{\nu}\nabla^{\nu}\nabla_{\mu}\chi_{\alpha_1 , \dots, \alpha_n}\right)\psi=\left(\nabla_{\nu}\nabla^{\nu}\nabla_{\mu}\chi_{\alpha_1 , \dots, \alpha_n}\right)\psi.\label{35}
\end{equation}
\begin{equation}
\nabla_{\nu}\psi(\nabla^{\nu}\nabla_{\mu}\chi_{\alpha_1 , \dots, \alpha_n})=i(c_{\nu}\nabla^{\nu}\nabla_{\mu}\chi_{\alpha_1 , \dots, \alpha_n})\psi+(z_{\nu}\nabla^{\nu}\nabla_{\mu}\chi_{\alpha_1 , \dots, \alpha_n})\psi.\label{36}
\end{equation}
With each term given in its expanded form in terms of the covariant derivatives of the path-ordered exponential, we substitute Eq. \ref{29}-\ref{36} into Eq. \ref{28} to obtain.
\begin{dmath}
\nabla_{\nu}\nabla^{\nu}\nabla_{\mu}\Psi_{\alpha_1,\dots,\alpha_n}=i(\nabla_{\nu}\nabla^{\nu}\chi_{\alpha_1 , \dots, \alpha_n})c_{\mu}\psi+(\nabla_{\nu}\nabla^{\nu}\chi_{\alpha_1 , \dots, \alpha_n})z_{\mu}\psi+i(\nabla^{\nu}\chi_{\alpha_1 , \dots, \alpha_n}\nabla_{\nu}c_{\mu})\psi-\nabla^{\nu}\chi_{\alpha_1 , \dots, \alpha_n} c_{\nu}c_{\mu}\psi+i(\nabla^{\nu}\chi_{\alpha_1 , \dots, \alpha_n} z_{\nu}c_{\mu})\psi+(\nabla^{\nu}\chi_{\alpha_1 , \dots, \alpha_n}\nabla_{\nu}z_{\mu})\psi+i(\nabla^{\nu}\chi_{\alpha_1 , \dots, \alpha_n} c_{\nu}z_{\mu})\psi+(\nabla^{\nu}\chi_{\alpha_1 , \dots, \alpha_n} z_{\nu}z_{\mu})\psi+i(\nabla_{\nu}\nabla^{\nu}c_{\mu})\Psi_{\alpha_1 , \dots, \alpha_n}-c_{\nu}\nabla^{\nu}c_{\mu}\Psi_{\alpha_1 , \dots, \alpha_n}+iz_{\nu}\nabla^{\nu}c_{\mu}\Psi_{\alpha_1 , \dots, \alpha_n}-\nabla^{\nu}(c^{\nu}c_{\mu})\Psi_{\alpha_1 , \dots, \alpha_n}-ic_{\nu}c^{\nu}c_{\mu}\Psi_{\alpha_1 , \dots, \alpha_n}-z_{\nu}c^{\nu}c_{\mu}\Psi_{\alpha_1 , \dots, \alpha_n}+i\nabla_{\nu}(z^{\nu}c_{\mu})\Psi_{\alpha_1 , \dots, \alpha_n}-c_{\nu}z^{\nu}c_{\mu}\Psi_{\alpha_1 , \dots, \alpha_n}+iz_{\nu}z^{\nu}c_{\mu}\Psi_{\alpha_1 , \dots, \alpha_n}+(\nabla_{\nu}\nabla^{\nu}z_{\mu})\Psi_{\alpha_1 , \dots,\alpha_n}+ic_{\nu}\nabla^{\nu}z_{\mu}\Psi_{\alpha_1,\dots,\alpha_n}+z_{\nu}\nabla^{\nu}z_{\mu}\Psi_{\alpha_1 , \dots, \alpha_n}+i\nabla_{\nu}(c^{\nu}z_{\mu})\Psi_{\alpha_1 , \dots, \alpha_n}-c_{\nu}c^{\nu}z_{\mu}\Psi_{\alpha_1 , \dots, \alpha_n}+iz_{\nu}c^{\nu}z_{\mu}\Psi_{\alpha_1 , \dots, \alpha_n}+\nabla_{\nu}(z^{\nu}z_{\mu})\Psi_{\alpha_1 , \dots, \alpha_n}+ic_{\nu}z^{\nu}z_{\mu}\Psi_{\alpha_1 , \dots,\alpha_n}+z_{\nu}z^{\nu}z_{\mu}\Psi_{\alpha_1 , \dots, \alpha_n}+i(\nabla_{\nu}\chi_{\alpha_1 , \dots, \alpha_n}\nabla^{\nu}c_{\mu})\psi-\nabla_{\nu}\chi_{\alpha_1 , \dots, \alpha_n} c^{\nu}c_{\mu}\psi+i(\nabla_{\nu}\chi_{\alpha_1 , \dots, \alpha_n} z^{\nu}c_{\mu})\psi+(\nabla_{\nu}\chi_{\alpha_1 , \dots, \alpha_n}\nabla^{\nu}z_{\mu})\psi+i(\nabla_{\nu}\chi_{\alpha_1 , \dots, \alpha_n} c^{\nu}z_{\mu})\psi+(\nabla_{\nu}\chi_{\alpha_1 , \dots, \alpha_n} z^{\nu}z_{\mu})\psi+i(\nabla_{\nu}c^{\nu}\nabla_{\mu}\chi_{\alpha_1 , \dots, \alpha_n})\psi-(c_{\nu}c^{\nu}\nabla_{\mu}\chi_{\alpha_1 , \dots,\alpha_n})\psi+i(z_{\nu}c^{\nu}\nabla_{\mu}\chi_{\alpha_1 , \dots, \alpha_n})\psi+(\nabla_{\nu}z^{\nu}\nabla_{\mu}\chi_{\alpha_1 , \dots,\alpha_n})\psi+i(c_{\nu}z^{\nu}\nabla_{\mu}\chi_{\alpha_1 , \dots, \alpha_n})\psi+(z_{\nu}z^{\nu}\nabla_{\mu}\chi_{\alpha_1 , \dots,\alpha_n})\psi+i(c^{\nu}\nabla_{\nu}\nabla_{\mu}\chi_{\alpha_1 , \dots, \alpha_n})\psi+(z^{\nu}\nabla_{\nu}\nabla_{\mu}\chi_{\alpha_1 , \dots,\alpha_n})\psi+\left(\nabla_{\nu}\nabla^{\nu}\nabla_{\mu}\chi_{\alpha_1 , \dots, \alpha_n}\right)\psi+i(c_{\nu}\nabla^{\nu}\nabla_{\mu}\chi_{\alpha_1 , \dots, \alpha_n})\psi+(z_{\nu}\nabla^{\nu}\nabla_{\mu}\chi_{\alpha_1 , \dots, \alpha_n})\psi.
\label{37}
\end{dmath}
We now make \(c_{\nu}c^{\nu}c_{\mu}\Psi_{\alpha_1 , \dots, \alpha_n}\) the subject in Eq. \ref{37} to yield;
\begin{dmath}
c_{\nu}c^{\nu}c_{\mu}\Psi_{\alpha_1 , \dots, \alpha_n}=i\left(\nabla_{\nu}\nabla^{\nu}\nabla_{\mu}\Psi_{\alpha_1 , \dots, \alpha_n}-i(\nabla_{\nu}\nabla^{\nu}\chi_{\alpha_1 , \dots, \alpha_n})c_{\mu}\psi-(\nabla_{\nu}\nabla^{\nu}\chi_{\alpha_1 , \dots, \alpha_n})z_{\mu}\psi
\\
-i(\nabla^{\nu}\chi_{\alpha_1 ,\dots,\alpha_n}\nabla_{\nu}c_{\mu})\psi+\nabla^{\nu}\chi_{\alpha_1 , \dots, \alpha_n} c_{\nu}c_{\mu}\psi-
i(\nabla^{\nu}\chi_{\alpha_1 , \dots, \alpha_n} z_{\nu}c_{\mu})\psi-
(\nabla^{\nu}\chi_{\alpha_1 , \dots, \alpha_n}\nabla_{\nu}z_{\mu})\psi-i(\nabla^{\nu}\chi_{\alpha_1 , \dots, \alpha_n} c_{\nu}z_{\mu})\psi-(\nabla^{\nu}\chi_{\alpha_1 , \dots, \alpha_n} z_{\nu}z_{\mu})\psi
\\
-i(\nabla_{\nu}\nabla^{\nu}c_{\mu})\Psi_{\alpha_1 , \dots, \alpha_n}+
c_{\nu}\nabla^{\nu}c_{\mu}\Psi_{\alpha_1 , \dots, \alpha_n}-iz_{\nu}\nabla^{\nu}c_{\mu}\Psi_{\alpha_1 , \dots, \alpha_n}+
\\
\nabla_{\nu}(c^{\nu}c_{\mu})\Psi_{\alpha_1 , \dots, \alpha_n}+
z_{\nu}c^{\nu}c_{\mu}\Psi_{\alpha_1 , \dots, \alpha_n}-i\nabla_{\nu}(z^{\nu}c_{\mu})\Psi_{\alpha_1 , \dots, \alpha_n}+
\\
c_{\nu}z^{\nu}c_{\mu}\Psi_{\alpha_1 , \dots, \alpha_n}-
iz_{\nu}z^{\nu}c_{\mu}\Psi_{\alpha_1 , \dots, \alpha_n}-(\nabla_{\nu}\nabla^{\nu}z_{\mu})\Psi_{\alpha_1 , \dots, \alpha_n}-
\\
ic_{\nu}\nabla^{\nu}z_{\mu}\Psi_{\alpha_1 , \dots, \alpha_n}-z_{\nu}\nabla^{\nu}z_{\mu}\Psi_{\alpha_1 , \dots, \alpha_n}-
i\nabla_{\nu}(c^{\nu}z_{\mu})\Psi_{\alpha_1 , \dots, \alpha_n}+
\\
c_{\nu}c^{\nu}z_{\mu}\Psi_{\alpha_1 , \dots, \alpha_n}-
iz_{\nu}c^{\nu}z_{\mu}\Psi_{\alpha_1 , \dots, \alpha_n}-\nabla_{\nu}(z^{\nu}z_{\mu})\Psi_{\alpha_1 , \dots, \alpha_n}-
\\
ic_{\nu}z^{\nu}z_{\mu}\Psi_{\alpha_1 , \dots, \alpha_n}-z_{\nu}z^{\nu}z_{\mu}\Psi_{\alpha_1 , \dots, \alpha_n}-i(\nabla_{\nu}\chi_{\alpha_1 , \dots, \alpha_n}\nabla^{\nu}c_{\mu})\psi+
\\
\nabla_{\nu}\chi_{\alpha_1 , \dots, \alpha_n} c^{\nu}c_{\mu}\psi-
i(\nabla_{\nu}\chi_{\alpha_1 , \dots, \alpha_n} z^{\nu}c_{\mu})\psi-(\nabla_{\nu}\chi_{\alpha_1 , \dots, \alpha_n}\nabla^{\nu}z_{\mu})\psi-
\\
i(\nabla_{\nu}\chi_{\alpha_1 , \dots, \alpha_n} c^{\nu}z_{\mu})\psi-(\nabla_{\nu}\chi_{\alpha_1 , \dots, \alpha_n} z^{\nu}z_{\mu})\psi-i(\nabla_{\nu}c^{\nu}\nabla_{\mu}\chi_{\alpha_1 , \dots, \alpha_n})\psi+
\\
(c_{\nu}c^{\nu}\nabla_{\mu}\chi_{\alpha_1 , \dots, \alpha_n})\psi-i(z_{\nu}c^{\nu}\nabla_{\mu}\chi_{\alpha_1 , \dots, \alpha_n})\psi-
(\nabla_{\nu}z^{\nu}\nabla_{\mu}\chi_{\alpha_1 , \dots, \alpha_n})\psi-
\\
i(c_{\nu}z^{\nu}\nabla_{\mu}\chi_{\alpha_1 , \dots, \alpha_n})\psi-(z_{\nu}z^{\nu}\nabla_{\mu}\chi_{\alpha_1 , \dots, \alpha_n})\psi-
i(c^{\nu}\nabla_{\nu}\nabla_{\mu}\chi_{\alpha_1 , \dots, \alpha_n})\psi-
\\
(z^{\nu}\nabla_{\nu}\nabla_{\mu}\chi_{\alpha_1 , \dots, \alpha_n})\psi-\left(\nabla_{\nu}\nabla^{\nu}\nabla_{\mu}\chi_{\alpha_1 , \dots, \alpha_n}\right)\psi-
i(c_{\nu}\nabla^{\nu}\nabla_{\mu}\chi_{\alpha_1 , \dots, \alpha_n})\psi
\\
-(z_{\nu}\nabla^{\nu}\nabla_{\mu}\chi_{\alpha_1 , \dots, \alpha_n})\psi\right).\label{38}
\end{dmath}
We may write
\begin{dmath}
c_{\nu}c^{\nu}c_{\mu}\Psi_{\alpha_1 , \dots, \alpha_n}=i((\nabla_{\nu}\nabla^{\nu}\nabla_{\mu}\psi)\chi_{\alpha_1 , \dots, \alpha_n}-i(\nabla_{\nu}\nabla^{\nu}c_{\mu})\Psi_{\alpha_1 , \dots, \alpha_n}+c_{\nu}\nabla^{\nu}c_{\mu}\Psi_{\alpha_1 , \dots, \alpha_n}-iz_{\nu}\nabla^{\nu}c_{\mu}\Psi_{\alpha_1 , \dots, \alpha_n}+\nabla_{\nu}(c^{\nu}c_{\mu})\Psi_{\alpha_1 , \dots, \alpha_n}+z_{\nu}c^{\nu}c_{\mu}\Psi_{\alpha_1 , \dots, \alpha_n}-i\nabla_{\nu}(z^{\nu}c_{\mu})\Psi_{\alpha_1 , \dots, \alpha_n}+c_{\nu}z^{\nu}c_{\mu}\Psi_{\alpha_1 , \dots, \alpha_n}-iz_{\nu}z^{\nu}c_{\mu}\Psi_{\alpha_1 , \dots, \alpha_n}-(\nabla_{\nu}\nabla^{\nu}z_{\mu})\Psi_{\alpha_1 , \dots, \alpha_n}-ic_{\nu}\nabla^{\nu}z_{\mu}\Psi_{\alpha_1 , \dots, \alpha_n}-z_{\nu}\nabla^{\nu}z_{\mu}\Psi_{\alpha_1 , \dots, \alpha_n}-i\nabla_{\nu}(c^{\nu}z_{\mu})\Psi_{\alpha_1 , \dots, \alpha_n}+c_{\nu}c^{\nu}z_{\mu}\Psi_{\alpha_1 , \dots, \alpha_n}-iz_{\nu}c^{\nu}z_{\mu}\Psi_{\alpha_1 , \dots, \alpha_n}-\nabla_{\nu}(z^{\nu}z_{\mu})\Psi_{\alpha_1 , \dots, \alpha_n}-ic_{\nu}z^{\nu}z_{\mu}\Psi_{\alpha_1 , \dots, \alpha_n}-z_{\nu}z^{\nu}z_{\mu}\Psi_{\alpha_1 , \dots, \alpha_n})\label{39}
\end{dmath}
This is another form, but it does not capture variations in the spinor field \(\chi_{\alpha_1 , \dots, \alpha_n}\).
Let us compact the terms in parentheses of Eq. \ref{38} to yield;
\begin{equation}
c_{\nu}c^{\nu}c_{\mu} \Psi_{\alpha_1 , \dots, \alpha_n}=iR_{\mu}\Psi_{\alpha_1 , \dots, \alpha_n}.\label{40}
\end{equation}
The imaginary unit \(i\) in Eq. \ref{40} introduces a profound feature in the QFWE by connecting the wavefunction's phase to the forces present in the dynamics of quantum systems. 
The term \(R_{\mu}\Psi_{\alpha_1 , \dots, \alpha_n}\) is;
\begin{dmath}
R_{\mu}\Psi_{\alpha_1 , \dots, \alpha_n}=\nabla_{\nu}\nabla^{\nu}\nabla_{\mu}\Psi_{\alpha_1 , \dots, \alpha_n}-i(\nabla_{\nu}\nabla^{\nu}\chi_{\alpha_1 , \dots, \alpha_n})c_{\mu}\psi-(\nabla_{\nu}\nabla^{\nu}\chi_{\alpha_1 , \dots, \alpha_n})z_{\mu}\psi
\\
-i(\nabla^{\nu}\chi_{\alpha_1 ,\dots,\alpha_n}\nabla_{\nu}c_{\mu})\psi+\nabla^{\nu}\chi_{\alpha_1 , \dots, \alpha_n} c_{\nu}c_{\mu}\psi-
i(\nabla^{\nu}\chi_{\alpha_1 , \dots, \alpha_n} z_{\nu}c_{\mu})\psi-
(\nabla^{\nu}\chi_{\alpha_1 , \dots, \alpha_n}\nabla_{\nu}z_{\mu})\psi-i(\nabla^{\nu}\chi_{\alpha_1 , \dots, \alpha_n} c_{\nu}z_{\mu})\psi-(\nabla^{\nu}\chi_{\alpha_1 , \dots, \alpha_n} z_{\nu}z_{\mu})\psi
\\
-i(\nabla_{\nu}\nabla^{\nu}c_{\mu})\Psi_{\alpha_1 , \dots, \alpha_n}+
c_{\nu}\nabla^{\nu}c_{\mu}\Psi_{\alpha_1 , \dots, \alpha_n}-iz_{\nu}\nabla^{\nu}c_{\mu}\Psi_{\alpha_1 , \dots, \alpha_n}+
\\
\nabla_{\nu}(c^{\nu}c_{\mu})\Psi_{\alpha_1 , \dots, \alpha_n}+
z_{\nu}c^{\nu}c_{\mu}\Psi_{\alpha_1 , \dots, \alpha_n}-i\nabla_{\nu}(z^{\nu}c_{\mu})\Psi_{\alpha_1 , \dots, \alpha_n}+
\\
c_{\nu}z^{\nu}c_{\mu}\Psi_{\alpha_1 , \dots, \alpha_n}-
iz_{\nu}z^{\nu}c_{\mu}\Psi_{\alpha_1 , \dots, \alpha_n}-(\nabla_{\nu}\nabla^{\nu}z_{\mu})\Psi_{\alpha_1 , \dots, \alpha_n}-
\\
ic_{\nu}\nabla^{\nu}z_{\mu}\Psi_{\alpha_1 , \dots, \alpha_n}-z_{\nu}\nabla^{\nu}z_{\mu}\Psi_{\alpha_1 , \dots, \alpha_n}-
i\nabla_{\nu}(c^{\nu}z_{\mu})\Psi_{\alpha_1 , \dots, \alpha_n}+
\\
c_{\nu}c^{\nu}z_{\mu}\Psi_{\alpha_1 , \dots, \alpha_n}-
iz_{\nu}c^{\nu}z_{\mu}\Psi_{\alpha_1 , \dots, \alpha_n}-\nabla_{\nu}(z^{\nu}z_{\mu})\Psi_{\alpha_1 , \dots, \alpha_n}-
\\
ic_{\nu}z^{\nu}z_{\mu}\Psi_{\alpha_1 , \dots, \alpha_n}-z_{\nu}z^{\nu}z_{\mu}\Psi_{\alpha_1 , \dots, \alpha_n}-i(\nabla_{\nu}\chi_{\alpha_1 , \dots, \alpha_n}\nabla^{\nu}c_{\mu})\psi+
\\
\nabla_{\nu}\chi_{\alpha_1 , \dots, \alpha_n} c^{\nu}c_{\mu}\psi-
i(\nabla_{\nu}\chi_{\alpha_1 , \dots, \alpha_n} z^{\nu}c_{\mu})\psi-(\nabla_{\nu}\chi_{\alpha_1 , \dots, \alpha_n}\nabla^{\nu}z_{\mu})\psi-
\\
i(\nabla_{\nu}\chi_{\alpha_1 , \dots, \alpha_n} c^{\nu}z_{\mu})\psi-(\nabla_{\nu}\chi_{\alpha_1 , \dots, \alpha_n} z^{\nu}z_{\mu})\psi-i(\nabla_{\nu}c^{\nu}\nabla_{\mu}\chi_{\alpha_1 , \dots, \alpha_n})\psi+
\\
(c_{\nu}c^{\nu}\nabla_{\mu}\chi_{\alpha_1 , \dots, \alpha_n})\psi-i(z_{\nu}c^{\nu}\nabla_{\mu}\chi_{\alpha_1 , \dots, \alpha_n})\psi-
(\nabla_{\nu}z^{\nu}\nabla_{\mu}\chi_{\alpha_1 , \dots, \alpha_n})\psi-
\\
i(c_{\nu}z^{\nu}\nabla_{\mu}\chi_{\alpha_1 , \dots, \alpha_n})\psi-(z_{\nu}z^{\nu}\nabla_{\mu}\chi_{\alpha_1 , \dots, \alpha_n})\psi-
i(c^{\nu}\nabla_{\nu}\nabla_{\mu}\chi_{\alpha_1 , \dots, \alpha_n})\psi-
\\
(z^{\nu}\nabla_{\nu}\nabla_{\mu}\chi_{\alpha_1 , \dots, \alpha_n})\psi-\left(\nabla_{\nu}\nabla^{\nu}\nabla_{\mu}\chi_{\alpha_1 , \dots, \alpha_n}\right)\psi-
i(c_{\nu}\nabla^{\nu}\nabla_{\mu}\chi_{\alpha_1 , \dots, \alpha_n})\psi
\\-(z_{\nu}\nabla^{\nu}\nabla_{\mu}\chi_{\alpha_1 , \dots, \alpha_n})\psi\label{41}
\end{dmath}
Eq. \ref{41} gives the equation for the GQM. Multiplying Eq. \ref{20} by \(\Psi_{\alpha_1 , \dots, \alpha_n}\) gives;
\begin{equation}
F_{\mu}\Psi_{\alpha_1 , \dots, \alpha_n}=\frac{\hbar^{2}}{2 \pi m}c_{\nu}c^{\nu}c_{\mu}\Psi_{\alpha_1 , \dots, \alpha_n}.\label{42}
\end{equation}
Substituting Eq. \ref{40} into Eq. \ref{41} yields;
\begin{equation}
F_{\mu}\Psi_{\alpha_1 , \dots, \alpha_n}=i\frac{\hbar^{2}}{2 \pi m}R_{\mu}\Psi_{\alpha_1 , \dots, \alpha_n}\label{43}
\end{equation}
Eq. \ref{43} is the \textbf{quantum force wave equation}.
\subsection{Establishing The Backreaction Framework}
The derived QFWE offers a generalized description of the evolution of quantum fields in response to their interactions with the environment, including curved spacetime. However, it does not account for the influence of quantum fields on the curvature itself. This yields a major limitation of the QFWE. Thus in this section, we derive a solution to this problem by modifying the Einstein's field equations with the QFWE and principles from quantum gravity to be able to address this problem.
First of all, we express general representation of an action in a four dimensional spacetime as:
\begin{equation}
S = \int d^4x \sqrt{-g} \mathcal{L},\label{44}
\end{equation}
where \(\sqrt{-g}\) is the determinant of the metric tensor, and \(\mathcal{L}\) is the Lagrangian density. To construct the Lagrangian which we denote as \(\mathcal{L_{\text{QF}}}\) for our framework, we incorporate quantum geometric principles motivated by loop quantum gravity (LQG).

In classical physics, pressure is defined as force per unit area. In the quantum regime, the area becomes quantized and the effective pressure naturally scales inversely with the quantized area eigenvalue. This is consistent with LQG, where discreteness in spacetime quantities is most prominently exhibited in the spectra of operators for area and volume \cite{11}. The area operator \(\mathcal{A}\) has eigenvalues given by:
\begin{equation}
\langle A \rangle = 8\pi \gamma l_{\text{Planck}}^2 \sqrt{j(j + 1)},\label{45}
\end{equation}
where \(\langle A \rangle\) is the area eigenvalue, \(\gamma\) is the Barbero-Immirzi parameter, \(l_{\text{Planck}}\) is the Planck length and \(j = 0, \frac{1}{2}, 1, \frac{3}{2}, \dots\) are irreducible representations of \(SU(2)\). The discreteness of the spectrum arises from the discrete nature of \(j\). For \(j = 0\), \(\langle A \rangle = 0\), corresponding to a nongeometric state, is excluded in this context. In the semi-classical limit (\(j \gg 1\)), \(\langle A \rangle\) becomes large, and the Lagrangian reduces to a form consistent with classical physics. For small \(j\), quantum corrections become significant.

Given the equivalence of the Lagrangian density to pressure, scaling \(\mathcal{L}\) inversely with \(\langle A \rangle\) ensures that the discreteness of the area spectrum is reflected in the Lagrangian. This motivates the form \(\mathcal{L} \propto \frac{1}{\langle A \rangle}\). To account for quantum forces, we include the term \(u_\mu F^\mu\), where \(F^\mu\) is the covariant quantum force which has energy density contributions represented by the time component \(\frac{E^2}{hc}\) and \(\mathbf{F}\), the spatial force component which represents forces exterted by a quantum field on spacetime. This term ensures consistency with the idea that quantum fields influence the geometry of spacetime. The field \(u_\mu\) is a unit covariant vector normalized as:
\[u_\mu u^\mu = \pm 1,\]
where \(+1\) corresponds to a timelike vector and \(-1\) to a spacelike vector. A timelike \(u_\mu\) defines a "preferred" flow of time, while a spacelike \(u_\mu\) represents a spatial direction, making \(u_\mu\) adaptable to the physical context of the system. The combination \(u_\mu F^\mu\) ensures the Lagrangian correctly incorporates the directionality of forces in spacetime.
From these considerations, the Lagrangian is constructed as:
\begin{equation}
\mathcal{L_{\text{QF}}} = \frac{1}{\langle A \rangle} u_\mu F^\mu,\label{46}
\end{equation}
and the corresponding action is:
\begin{equation}
S_{\text{QF}} = \int d^4x \sqrt{-g} \frac{1}{\langle A \rangle} u_\mu F^\mu.\label{47}
\end{equation}
This action explicitly incorporates the quantum geometry of spacetime through the discrete area operator \(\langle A \rangle\) and the contributions of quantum forces, making it consistent with LQG principles. 
With our constructed action, we derive a new field equation which is a modification of the Einstein's field equations by applying the principle of stationary action writing it as the functional derivative of the actions with respect to the metric \(g^{\mu\nu}\). 
\begin{equation}
\frac{\delta S_{\text{EH}}}{\delta g^{\mu\nu}}+\frac{\delta S_{\text{QF}}}{\delta g^{\mu\nu}}=0,\label{48}
\end{equation}
where \(S_{\text{EH}}\) is the Einstein-hilbert action. We make \(F^{\mu}\) explicitly dependent on the metric tensor by computing \(F^{\mu}=F_{\nu}g^{\mu\nu}\). However, \(\langle A \rangle\) and \(u_{\nu}\) are taken not explicitly dependent on the metric tensor. With these considerations, we then take the variation of the actions to yield:
\begin{equation}
\frac{\sqrt{-g}c^{4}}{16\pi G}G_{\mu\nu}+\frac{\sqrt{-g}}{\langle A \rangle}(u_{\mu}F_{\nu}-\frac{1}{2}g_{\mu\nu}u_{\rho}F^{\rho})=0.\label{49}
\end{equation}
Where \(G_{\mu\nu}\) is the Einstein's tensor. We further make the Einstein's tensor the subject to yield
\begin{equation}
G_{\mu\nu}=\frac{8\pi G}{\langle A \rangle  c^{4}}(g_{\mu\nu}u_{\rho}F^{\rho}-2u_{\mu}F_{\nu})\label{50}
\end{equation}
We now embed the QFWE into Eq. \ref{50}. To achieve this, we multiply through by the wave function \(\Psi_{\alpha_1 ,\dots,\alpha_n}\) to yield
\begin{equation}
G_{\mu\nu}\Psi_{\alpha_1 ,\dots,\alpha_n}=\frac{8\pi G}{\langle A \rangle  c^{4}}(g_{\mu\nu}u_{\rho}F^{\rho}\Psi_{\alpha_1 ,\dots,\alpha_n}-2u_{\mu}F_{\nu}\Psi_{\alpha_1 ,\dots,\alpha_n}).\label{51}
\end{equation}
Substituting Eq. \ref{43} yields
\begin{equation}
G_{\mu\nu}\Psi_{\alpha_1 ,\dots,\alpha_n}=\frac{8\pi G}{\langle A \rangle  c^{4}}\left(g_{\mu\nu}u_{\rho}\left(i\frac{\hbar^{2}}{2 \pi m}R^{\rho}\Psi_{\alpha_1 , \dots, \alpha_n}\right)-2u_{\mu}\left(i\frac{\hbar^{2}}{2 \pi m}R_{\nu}\Psi_{\alpha_1 , \dots, \alpha_n}\right)\right),\label{52}
\end{equation}
\begin{equation}
G_{\mu\nu}\Psi_{\alpha_1 ,\dots,\alpha_n}=i\frac{\hbar^{2}}{2 \pi m}\frac{8\pi G}{\langle A \rangle  c^{4}}\left(g_{\mu\nu}u_{\rho}R^{\rho}-2u_{\mu}R_{\nu}\right)\Psi_{\alpha_1 , \dots, \alpha_n}.\label{53}
\end{equation}
We rewrite and substitute the area operator eigenvalues as \(\langle A \rangle=8\pi l_{Planck}^{2}\langle q \rangle\) where \(\langle q \rangle = \gamma\sqrt{j(j+1)}\). We also replace the terms in the bracket by \(M_{\mu\nu}\) to obtain
\begin{equation}
G_{\mu\nu}\Psi_{\alpha_1 ,\dots,\alpha_n}=i\frac{\hbar}{2\pi mc\langle q\rangle}M_{\mu\nu}\Psi_{\alpha_1 ,\dots,\alpha_n}.\label{54}
\end{equation}
Eq. \ref{54} brings us to our solution to the backreaction effect of quantum fields to the structure of spacetime.  It follows that the equation has many solutions. However, a detailed study of the solutions and viability of the equation is beyond the scope of this study and will be the subject of future research.

\section{Physical Interpretations And Discussion}\label{4}
The QFWE, developed to describe quantum forces in curved spacetime, represents a significant shift in our understanding of the fundamental interactions between quantum particles and their environments. This provides a universal approach for understanding how quantum forces interact both locally and non-locally with all massive quantum particles with various spin representations in all spacetime geometries. In the following subsections we discuss the general significance of the QFWE.
\subsection{New Definition of Quantum Forces}
A central insight into the QFWE is the redefinition of force within quantum theory. In conventional quantum mechanics, forces are typically derived from interaction potentials, such as the electromagnetic potential in the gauge theory. However, in the QFWE, forces emerge from the interaction between the wavefunction of a particle and the local spacetime curvature. This perspective implies that quantum forces are inherently sensitive to both the geometric and dynamic properties of spacetime, fundamentally distinguishing them from the classical and standard quantum forces. The generalized quantum force \(F_{\mu}\), which includes an energy-dependent time component \(E^{2}/hc\), suggests that the quantum forces vary with the relativistic energy of the particle. Physically, this means that highly energetic particles, such as those in extreme astrophysical environments or high-energy experiments, can experience stronger or qualitatively different quantum forces than low-energy particles. This energy-dependence directly ties quantum force behavior to spacetime distortions caused by gravitational fields, making the QFWE particularly relevant for studying quantum particle dynamics in extreme settings such as black holes or the early universe.
The embedding of covariant derivatives and spin connections in the QFWE also ensures that the quantum force accounts for the spacetime curvature and particle spin. For instance, the inclusion of spin connection terms highlights the interplay between the intrinsic angular momentum of quantum particles and spacetime geometry, which can manifest as spin-dependent quantum forces in rotating spacetimes, such as Kerr black holes. This spin-curvature coupling is a key feature that distinguishes the QFWE from standard quantum mechanical frameworks and opens new avenues for studying spin effects in quantum gravity. Furthermore, the framework’s capacity to incorporate higher-order curvature terms, such as those associated with scalar degrees of freedom such as the scalaron \cite{12}, suggests that it can describe additional gravitational effects predicted by extensions of general relativity, such as f(R) theories or string-inspired models.
\subsection{Unification of Quantum Forces}
Over the past few decades, the role of torsion in gravity has been extensively investigated, particularly in efforts to bring gravity closer to a gauge formulation and incorporate spin within a geometric framework \cite{13,14,15}. The QFWE represents an attempt to unify these interactions under a single mathematical structure. According to this framework, a particle should exist on which gauge forces act more strongly than gravity and scalar forces combined \cite{16}, and the QFWE may provide a means for predicting such a particle.
The electromagnetic force, as described by quantum electrodynamics (QED), can be incorporated into the QFWE through the gauge field \(A^{a}_{\mu}\) embedded in the field-coupling vector \(c_{\mu}\). The photon field, coupled to the charge of a particle via the coupling constant \(e\), allows the QFWE to describe electromagnetic interactions in a gauge-invariant manner which yields physically real results according to the nearly universally accepted claim that only gauge-invariant quantities can be physically real \cite{17}. The path-ordered exponential \(\psi\) in the wavefunction incorporates electromagnetic phase factors, enabling effects such as the Aharonov-Bohm effect, where particles are influenced by the vector potential even in regions where the electromagnetic field strength is zero.
The weak nuclear force, mediated by \(W\) and \(Z\) bosons \cite{18,19,20}, which drives processes such as beta decay, can also be described within this framework by including the \(SU(2)\) gauge fields. With appropriate gauge field terms in the field-coupling vector, the QFWE can model how particles couple to weak interactions, particularly for fermions (such as electrons and neutrinos) involved in weak processes. Similarly, the \(SU(3)_{C}\) gauge fields \(G^{a}_{\mu}\) \cite{21} can be incorporated within \(g^{a}A^{a}_{\mu}T^{a}\), where \(A^{a}_{\mu}\) includes the gluon fields and \(T^{a}\) represents the generators of the \(SU(3)\) gauge group. This extension allows the QFWE to capture the interactions between quarks mediated by gluons. Because the QFWE includes gauge invariance and path-dependent effects, it can describe non-Abelian gauge theories such as quantum chromodynamics (QCD), where gluons interact with each other. This may enable the QFWE to account for phenomena such as confinement, where quarks are bound together within hadrons.
Consequently, the QFWE appears capable of handling all gauge interactions described by the gauge group \(SU(3)_{C} \times SU(2)_{L} \times U(1)_{Y}\) \cite{22}. Furthermore, if dark matter is associated with a new gauge field—sometimes referred to as a “dark photon" or a particle from a hidden sector—the QFWE could potentially model this interaction through additional gauge terms in the field-coupling vector \(c_{\mu}\). Similarly, hypothetical forces related to dark energy, such as “dark gravity" \cite{23,24,25}, can be incorporated if mediated by a field that couples to spacetime or Standard Model particles.
One of the unique features of the QFWE is its integration of the spacetime curvature and gravitational effects through the spin connection \(\omega^{ab}_{\mu}\) in the field-coupling vector. The GQM \(R_{\mu}\) in the equation includes covariant derivatives that account for spacetime curvature, allowing the wavefunctions to interact with gravitational fields in a manner compatible with general relativity. Through the spin connection term \(\omega^{ab}_{\mu} J_{ab}\), the QFWE enables particles with spin to couple to spacetime curvature, potentially describing phenomena such as spin precession in gravitational fields and providing a quantum framework for studying the influence of gravity on quantum states.
Importantly, the QFWE treats gravity as a quantum force acting on quantum particles, rather than requiring full quantization of spacetime itself. Gravity can exist without mass \cite{26}, which makes it possible for the QFWE to fit well in describing gravitational effects on quantum systems on an equal footing with other fundamental forces. This semi-classical approach is particularly effective in describing confined quantum particles in the presence of a strong spacetime curvature or other force fields. For example, particles bound within gauge fields, such as quarks confined within hadrons by the strong forces, experience quantized interactions that are seamlessly modeled within the QFWE. Likewise, particles confined by spacetime curvature, such as those near massive celestial objects such as black holes, are subject to gravitational forces mediated by the spin connection and curvature terms within the QFWE. The QFWE framework also extends to non-confined particles. For instance, free particles propagating through weak gravitational fields or interacting with minimal gauge fields can be described accurately because the QFWE accounts for the phase shifts and quantum effects induced by such fields. This broad applicability highlights the utility of the QFWE in describing both confined and non-confined quantum systems and can further redefine the meaning of quantum gravity by quantizing gravitational force fields instead of quantizing spacetime itself.
By quantizing all fundamental forces while incorporating the geometric influence of the spacetime curvature, the QFWE unifies particle interactions under a consistent and mathematically rigorous framework. This unification spans the Standard Model forces, gravitational interactions, and potentially beyond-standard model phenomena, such as dark matter and dark energy.
This unified approach suggests that the QFWE could serve as a \textit{generalized wave equation for all fundamental interactions}, allowing particles of various spins and charges to experience the combined effects of gauge and gravitational fields \cite{27}.

\subsection{Relevance to Quantum Particle Dynamics in Spacetime}
The QFWE provides a unified description of quantum forces in diverse spacetime geometries, from flat to highly curved regions. For example, in flat spacetime, the QFWE reduces to standard quantum mechanical expressions, ensuring consistency with existing quantum theory within the appropriate limit. However, in strongly curved spacetimes, such as those near neutron stars or within the event horizons of black holes, the QFWE introduces corrections to quantum dynamics that are absent in conventional quantum mechanics. These corrections can manifest as measurable effects in phenomena such as gravitational time dilation or tidal forces at quantum scales. For instance, quantum particles propagating in a rapidly expanding spacetime, such as during cosmic inflation, can experience phase shifts or quantum forces directly attributable to the spacetime curvature. Such dynamics are inaccessible to traditional quantum field theory in curved spacetime, which treats particles as excitations of fields on a fixed background and does not account for the dynamical coupling between the wave function and spacetime geometry.
One of the most striking implications of the QFWE is the coupling vector \(c_{\mu}\), which encapsulates the interaction between quantum particles, gauge fields, and spacetime curvature. This vector generalizes the concept of force by incorporating both gauge symmetries and spacetime dynamics, enabling the QFWE to account for effects such as electromagnetic-gravitational coupling or quantum corrections to gravitational lensing. By expanding \(c_{\mu}\) to include hypothetical interactions, such as those arising from dark energy or additional dimensions, the QFWE could potentially provide a framework for studying BSM physics in both quantum and gravitational contexts. The inclusion of gauge fields in \(c_{\mu}\) also establishes a natural connection to gauge-gravity dualities \cite{28}, suggesting that the QFWE might serve as a bridge between quantum gravity approaches based on holographic principles and those grounded in traditional field theory.
\subsection{Backreaction Effect and Quantum Spacetime}
At its core, the modified Einstein field equation (Eq . \ref{54} ) expresses a profound physical truth: spacetime is not a static stage on which quantum fields perform, but a dynamic, malleable entity that responds directly to the quantum behavior of matter. In this framework, curvature is no longer dictated by classical energy and momentum alone, but by the full, living structure of the quantum wavefunction itself, its internal spin, its gauge interactions, its coherence, and its nonlocal entanglements.

The backreaction encoded in this equation is not passive.  Rather it suggests that curvature itself may oscillate dynamically due to interactions with quantum matter. Spacetime does not merely adjust itself to the average energy of quantum fields, it resonates in response to how quantum states evolve, how they interfere, and how they are entangled across regions. This means that every fluctuation, every correlation, every unit of quantum information plays a role in sculpting the geometry of the universe. Unlike classical general relativity, where curvature is a smooth, continuous deformation of spacetime in response to energy and momentum.

The presence of the imaginary unit \(i\) suggests that curvature does not merely adjust statically to quantum matter but instead exhibits oscillatory behavior, akin to quantum wave dynamics. This implies that spacetime, at the quantum level, may not be a rigid geometric construct but a dynamical quantum medium capable of undergoing harmonic variations.

A key element of this oscillation lies in the dimensionless operator \(\langle q \rangle\) where \(j\) represents discrete quantum states of the \(SU(2)\) group, encoding the fundamental quantization of spacetime. This discrete nature implies that the curvature does not shift continuously, but rather undergoes quantized oscillatory transitions, driven by interactions with quantum wavefunctions.

Physically, this means that gravitational fields fluctuate in response to quantum states, much like electromagnetic fields oscillate in response to charged particles. The tensor \(M_{\mu\nu}\) containing the GQM further refines this interaction, incorporating higher-order derivatives that generate curvature waves, propagating through spacetime much like quantum mechanical wavefunctions evolve over time.

Also the equation tells us that when a quantum field exists in a given region, its presence is felt not just through energy density, but through how it moves, how it couples to force carriers, and how it remembers its own history. The structure of the field, its spin content, its gauge charge, and its path through spacetime, feeds back into the very curvature it travels through. The universe, then, becomes a dialogue between quantum matter and geometry: the field evolves in a curved background, but that curvature is itself being reshaped by the quantum state.

Even more striking is the influence of nonlocality. Quantum fields, especially when entangled, generate effects that cannot be confined to a point or a localized region. This theory captures that insight directly. Spacetime responds not only to what's happening here and now, but also to what's happening there and elsewhere. Through nonlocal corrections, distant correlations weave themselves into the fabric of curvature, as if spacetime is sensitive to the global tapestry of quantum relationships.

This leads to a new picture of gravity, not just as a force but as an emerging property of the quantum informational structure. Spacetime is no longer the backdrop; it is a byproduct of quantum processes. The wavefunction doesn’t just live on spacetime, it dynamically modifies and sustains it. In this light, black holes, cosmological expansion, and the flow of time itself are all natural outcomes of the quantum backreaction mechanism.

Ultimately, the equation presents a radically new understanding of spacetime: a dynamic, oscillatory fabric that resonates with quantum matter, blurring the distinction between geometry and quantum fields. Gravity is not merely influenced by matter, it is the shadow cast by quantum reality onto the continuum of space and time. If validated, this would redefine gravity, not as a classical force, but as an emergent quantum phenomenon shaped by oscillating spacetime curvature.

\subsection{Comparison with Existing Theories}
This framework offers a fundamentally different perspective compared to existing frameworks like quantum field theory in curved spacetime and approaches in quantum gravity such as loop quantum gravity and string theory. Its unique features provide both conceptual and practical advantages for studying the interaction of quantum particles with spacetime geometry. 

\subsubsection{Quantum Field Theory in Curved Spacetime}
Quantum field theory in curved spacetime assumes a classical, fixed spacetime background on which quantum fields propagate. The effects of spacetime curvature manifest as modifications to particle propagation, such as redshifts or changes in dispersion relations, but the framework does not allow for the dynamic coupling of quantum fields to the curvature itself. In this context, spacetime acts purely as a passive stage for quantum phenomena. The established framework, in contrast, treats the wave function as an active participant in spacetime dynamics, establishing a bidirectional relationship between quantum particles and the geometry of spacetime. This interaction is explicitly captured through \(c_{\mu}\), which links gauge fields, spin effects, and curvature to the quantum dynamics of particles. By incorporating covariant derivatives and spin-connection terms into the wavefunction evolution, the QFWE reflects how spacetime curvature directly influences—and is influenced by—quantum particles, aligning more closely with the goals of quantum gravity.

\subsubsection{Loop Quantum Gravity and String Theory}
The QFWE also contrasts with leading quantum gravity theories, such as loop quantum gravity and string theory, in its scope and approach. Loop quantum gravity focuses on the quantization of spacetime itself, treating spacetime geometry as a discrete quantum entity \cite{29,30,31}. Although powerful, it does not directly address the behavior of quantum particles or forces within a curved spacetime continuum. In contrast, the QFWE operates within the continuous, four-dimensional framework of general relativity and provides a detailed description of how quantum forces arise from the interaction between wave functions and curved spacetime. This makes it more accessible for modeling particle-level phenomena in curved spacetimes, while maintaining compatibility with established physics.

String theory, on the other hand, posits that particles and forces emerge as excitations of one-dimensional strings propagating in higher-dimensional spacetimes \cite{32}.  String theory has the potential to unify all fundamental forces; its reliance on higher dimensions and a significantly more abstract formalism make it computationally intensive and less immediately applicable to four-dimensional physical systems. The QFWE, by remaining within the traditional four-dimensional spacetime framework, offers a computationally tractable and conceptually straightforward tool for studying quantum forces, making it more practical for a range of applications in high-energy physics and astrophysics.
\subsubsection{Unique Position of the QFWE}
The QFWE occupies a distinctive niche by combining the best elements of existing approaches while addressing their limitations. Unlike quantum field theory in curved spacetime and semi-classical gravity, it establishes a dynamic interaction between wave functions and spacetime geometry, providing a direct link between quantum states and spacetime curvature. Unlike loop quantum gravity, it does not require a discrete spacetime structure, and unlike string theory, it does not rely on higher dimensions or abstract string dynamics. Instead, the QFWE leverages established concepts—such as covariant derivatives, spin connections, and gauge interactions—within the familiar framework of general relativity, while extending these ideas to incorporate quantum forces as emergent phenomena.

\subsection{Symmetry Considerations and Energy Conservations}
The QFWE was constructed to preserve key physical symmetries, ensuring consistency with the fundamental principles of physics. First, the covariant structure of the equation guarantees Lorentz invariance, because it is explicitly formulated to respect the local symmetry of spacetime under transformations between inertial observers. Similarly, gauge invariance is maintained through the inclusion of gauge field couplings in \(c_{\mu}\), which ensures that quantum forces transform appropriately under local gauge transformations. This feature allows the QFWE to integrate seamlessly with established gauge theories, such as quantum electrodynamics or the Standard Model, while accommodating extensions that involve additional gauge fields or symmetries.
The energy variation in the QFWE is more subtle, as the framework inherently incorporates dynamical spacetime effects, which can lead to energy transfer from the quantum system to the spacetime geometry, and vice versa. In dynamic spacetimes, the QFWE predicts energy transfer processes that can offer new insights into phenomena such as Hawking radiation or the quantum effects of cosmic expansion. These processes could provide novel tests of the QFWE in both theoretical and experimental settings.
\subsection{Elements Of Quantum Forces}
Due to the lengthy structure of the GQM tied to the wavefunction \(R_{\mu}\Psi_{\alpha_1 ,\dots,\alpha_n}\), we classify the terms based on their fundamental role in describing the fundamental forces of nature. In order for us to better understand these elements of quantum forces, we give them names to represent their fundamental nature. From these terms, we can obtain four main categories: foundational force, grand local force, grand non-local force and grand complementary force. Below, we discuss these separate elements of quantum force and the equations that represent each of them in the QFWE.
\subsubsection{Foundational Force}
The foundational force is described by the equation
\begin{equation}
F_\mu^{(1)} \Psi_{\alpha_1, \dots, \alpha_n} = i \frac{\hbar^2}{2\pi m} R_\mu^{(1)} \Psi_{\alpha_1, \dots, \alpha_n},
\end{equation}
\begin{equation}
R_\mu^{(1)} \Psi_{\alpha_1 , \dots, \alpha_n} = \nabla_\nu \nabla^\nu \nabla_\mu\Psi_{\alpha_1 , \dots, \alpha_n}  - (\nabla_\nu \nabla^\nu \nabla_\mu \chi_{\alpha_1 , \dots, \alpha_n}) \psi.
\end{equation}
This element of quantum force represents the most fundamental level of quantum interactions, originating directly from fluctuations of the wave function \(\Psi_{\alpha_1, \dots, \alpha_n}\). It serves as the source of all the forces in quantum systems and provides the mechanical force that drives quantum dynamics. The GQM for this elementary force \(R_\mu^{(1)}\) reflects the third-order derivative of the wave function and the tensor-spinor field, describing the intrinsic quantum behaviors. This force emerges purely from the self-consistent dynamics of the wave function. It is neither explicitly local or non-local, but rather exists as the bedrock of quantum systems, from which all forces arise. Physically, it represents the quantum vacuum energy and baseline fluctuations inherent to any quantum state. This is especially significant in explaining phenomena such as zero-point energy, which arises from the ever-present fluctuations of the quantum vacuum. It also underpins the dynamics of quantum fields that seed the structure of the universe during cosmic inflation. By describing quantum systems in their most basic form, \(F_\mu^{(1)}\) lays the foundation for understanding how higher-level interactions emerge from primordial quantum backgrounds.
\subsubsection{Grand Local Force}
The grand local force is described by the equation 
\begin{equation}
F_\mu^{(2)} \Psi_{\alpha_1, \dots, \alpha_n} = i \frac{\hbar^2}{2\pi m} R_\mu^{(2)} \Psi_{\alpha_1, \dots, \alpha_n},
\end{equation}
\begin{dmath}
R^{(2)}_{\mu}\Psi_{\alpha_1 , \dots, \alpha_n}=\nabla^{\nu}\chi_{\alpha_1 , \dots, \alpha_n} c_{\nu}c_{\mu}\psi+c_{\nu}\nabla^{\nu}c_{\mu}\Psi_{\alpha_1 , \dots, \alpha_n}+\nabla_{\nu}(c^{\nu}c_{\mu}) \Psi_{\alpha_1 , \dots, \alpha_n}+
\\
\nabla_{\nu}\chi_{\alpha_1 , \dots, \alpha_n} c^{\nu}c_{\mu}\psi+(c_{\nu}c^{\nu}\nabla_{\mu}\chi_{\alpha_1 , \dots, \alpha_n})\psi+\nabla^{\nu}c_{\nu}c_{\mu}\Psi_{\alpha_1 , \dots, \alpha_n}-
\\
i(\nabla_{\nu}\nabla^{\nu}\chi_{\alpha_1 , \dots, \alpha_n})c_{\mu}\psi-i(\nabla^{\nu}\chi_{\alpha_1 , \dots, \alpha_n} \nabla_{\nu}c_{\mu})\psi-
i(\nabla_{\nu}\chi_{\alpha_1 , \dots, \alpha_n} \nabla^{\nu}c_{\mu})\psi-
\\
i(\nabla_{\nu}c^{\nu}\nabla_{\mu}\chi_{\alpha_1 , \dots, \alpha_n})\psi-i(c^{\nu}\nabla_{\nu}\nabla_{\mu}\chi_{\alpha_1 , \dots, \alpha_n})\psi-i(c_{\nu}\nabla^{\nu}\nabla_{\mu}\chi_{\alpha_1 , \dots, \alpha_n})\psi.
\end{dmath}
This element of quantum force describes the forces that arise from localized interactions. This is the force with which we are most familiar in describing the fundamental forces of nature. It operates within specific regions of spacetime and is mediated by fields. The structure of \(R_\mu^{(2)}\) highlights the role of gauge symmetries in quantum systems. It contains all terms that incorporate only \(c_{\mu}\), reflecting how the particles interact with their environment through local field configurations. The inclusion of the spin-connection term and covariant derivatives makes it compatible with spacetime curvature and quantum-mechanical principles. 

The grand local force is fundamentally spacetime-local in nature. However, it allows for apparent non-locality as an emergent property of collective local dynamics like the term \(\nabla_{\nu}c^{\nu}c_{\mu} \Psi_{\alpha_1, \dots, \alpha_n}\) which couples the divergence of \(c^{\nu}\) with its \(\mu\)-component. This term implies that the divergence of the field-coupling vector in one direction can influence quantum forces in another, emphasizing the interconnected and multidirectional nature of quantum field interactions as an emergent non-locality from local dynamics in curved spacetime. Physically, this force underpins the standard model and the forces that govern matter and energy in everyday experiences.
\subsubsection{Grand Non-local Force}
The grand non-local force is defined by
\begin{equation}
F_\mu^{(3)} \Psi_{\alpha_1, \dots, \alpha_n} = i \frac{\hbar^2}{2\pi m} R_\mu^{(3)} \Psi_{\alpha_1, \dots, \alpha_n},
\end{equation}
\begin{dmath}
R^{(3)}_{\mu}\Psi_{\alpha_1 , \dots, \alpha_n}=-(\nabla_{\nu}\nabla^{\nu}\chi_{\alpha_1 , \dots, \alpha_n})z_{\mu}\psi-(\nabla^{\nu}\chi_{\alpha_1 , \dots, \alpha_n}\nabla_{\nu}z_{\mu})\psi-(\nabla^{\nu}\chi_{\alpha_1 , \dots, \alpha_n} z_{\nu}z_{\mu})\psi-(\nabla_{\nu}\nabla^{\nu}z_{\mu})\Psi_{\alpha_1 , \dots, \alpha_n}-z_{\nu}\nabla^{\nu}z_{\mu}\Psi_{\alpha_1 , \dots, \alpha_n}-\nabla_{\nu}(z^{\nu}z_{\mu})\Psi_{\alpha_1 , \dots, \alpha_n}-z_{\nu}z^{\nu}z_{\mu}\Psi_{\alpha_1 , \dots, \alpha_n}-(\nabla_{\nu}\chi_{\alpha_1 , \dots, \alpha_n} \nabla^{\nu}z_{\mu})\psi-\nabla_{\nu}\chi_{\alpha_1 , \dots, \alpha_n} z^{\nu}z_{\mu}\psi-(\nabla_{\nu}z^{\nu}\nabla_{\mu}\chi_{\alpha_1 , \dots, \alpha_n})\psi-(z_{\nu}z^{\nu}\nabla_{\mu}\chi_{\alpha_1 , \dots, \alpha_n})\psi-(z^{\nu}\nabla_{\nu}\nabla_{\mu}\chi_{\alpha_1 , \dots, \alpha_n})\psi-(z_{\nu}\nabla^{\nu}\nabla_{\mu}\chi_{\alpha_1 , \dots, \alpha_n})\psi.
\end{dmath}
This element of quantum force  fundamentally governs non-local interactions, where particles influence one another across spacetime without being constrained by locality. Unlike grand local force, this force is inherently non-local and transcends classical notions of causality and locality. The non-local nature of this elementary force is encoded in the \(z_{\mu}\) corrections,  which involve higher-order commutators of \(c_{\mu}\) along the paths in spacetime. These corrections describe the interaction of quantum particles across multiple spacetime points, allowing for long-range correlations and influences.

The grand non-local force is particularly important in quantum entanglement, which lies at the heart of many quantum phenomena. For example, in the Einstein-Podolsky-Rosen (EPR) paradox, two entangled  quantum systems exhibit correlated behaviors even when separated by large distances \cite{33}. This elementary quantum force shows how all known fundamental forces of nature can non-locally interact with all quantum particles. This makes it significant in foundational quantum mechanics and has implications for quantum gravity. It also provides a natural way of modeling holographic dualities, such as the AdS/CFT correspondence, where bulk spacetime phenomena are encoded in non-local boundary theories.
\subsubsection{Grand Complimentary Force}
The grand complementary force is represented by
\begin{equation}
F_\mu^{(4)} \Psi_{\alpha_1, \dots, \alpha_n} = i \frac{\hbar^2}{2\pi m} R_\mu^{(4)} \Psi_{\alpha_1, \dots, \alpha_n},
\end{equation}
\begin{dmath}
R^{(4)}_{\mu}\Psi_{\alpha_1 , \dots, \alpha_n}=c_{\nu}c^{\nu}z_{\mu}\Psi_{\alpha_1 , \dots, \alpha_n}+c_{\nu}z^{\nu}c_{\mu}\Psi_{\alpha_1 , \dots, \alpha_n}+z_{\nu}c^{\nu}c_{\mu}\Psi_{\alpha_1 , \dots, \alpha_n}-
\\
i(\nabla^{\nu}\chi_{\alpha_1 , \dots, \alpha_n} z_{\nu}c_{\mu})\psi-i(\nabla^{\nu}\chi_{\alpha_1 , \dots, \alpha_n} c_{\nu}z_{\mu})\psi-iz_{\nu}\nabla^{\nu}c_{\mu}\Psi_{\alpha_1 , \dots, \alpha_n}-
\\
i\nabla_{\nu}(z^{\nu}c_{\mu})\Psi_{\alpha_1 , \dots, \alpha_n}-iz_{\nu}z^{\nu}c_{\mu}\Psi_{\alpha_1 , \dots, \alpha_n}-ic_{\nu}z^{\nu}z_{\mu}\Psi_{\alpha_1 , \dots, \alpha_n}-
\\
i(\nabla_{\nu}\chi_{\alpha_1 , \dots, \alpha_n} z^{\nu}c_{\mu})\psi-
i(\nabla_{\nu}\chi_{\alpha_1 , \dots, \alpha_n} c^{\nu}z_{\mu})\psi-i(z_{\nu}c^{\nu}\nabla_{\mu}\chi_{\alpha_1 , \dots, \alpha_n})\psi-
\\
i(c_{\nu}z^{\nu}\nabla_{\mu}\chi_{\alpha_1 , \dots, \alpha_n})\psi-
ic_{\nu}\nabla^{\nu}z_{\mu}\Psi_{\alpha_1 , \dots, \alpha_n}-i\nabla_{\nu}(c^{\nu}z_{\mu})\Psi_{\alpha_1 , \dots, \alpha_n}-
\\
iz_{\nu}c^{\nu}z_{\mu}\Psi_{\alpha_1 , \dots, \alpha_n}.
\end{dmath}
This elementary quantum force describes the interplay and coexistence of locality and non-locality. Unlike previous elements of quantum force, this force treats locality and non-locality as complimentary aspects of the same physical reality. It captures the dual nature of quantum systems, where local and non-local systems work together to shape quantum dynamics. This force heavily relies on \(c_{\mu}\) capturing local dynamics and \(z_{\mu}\) capturing non-local dynamics. The interaction between these terms highlights how local and non-local processes can give rise to each other. For example, complementary forces could explain how the information encoded in the black hole's event horizon is simultaneously correlated with the hawking radiation emitted far away.

\section{Potential Applications}\label{5}
\subsection{Black Hole Physics}
Black holes are the most compact objects in the universe and provide probes for the strongest gravitational fields \cite{34}. These intense gravitational fields, along with their interactions with quantum particles, can be studied using the QFWE. The QFWE can help physicists gain insights into the formation of a black hole shadow, which arises because of the strong deflection of light rays near the black hole and the absorption of light by the event horizon \cite{35,36,37}. Studying this shadow can shed light on the behavior of matter and radiation under extreme conditions, which is crucial for understanding phenomena such as Hawking radiation—particularly as it applies to Dirac particles \cite{38}—and black hole thermodynamics, given the close relationship between gravitational dynamics and horizon thermodynamics \cite{39,40,41}. In these regions, quantum effects and spacetime curvature interact in complex ways. For instance, the QFWE can help describe particle dynamics near the event horizon, offering deeper insights into Hawking radiation and the mechanisms by which particles may escape from black holes through quantum processes.
\subsection{Quantum Information Theory}
The QFWE presents novel opportunities in quantum information theory by leveraging its gauge invariance and non-local corrections to address key challenges in quantum computing and information preservation. Gauge invariance, a fundamental property of the QFWE, provides a natural mechanism for constructing gauge-protected quantum states and error-correcting codes that are resilient to environmental disturbances and decoherence. This property is especially valuable for fault-tolerant quantum computing in systems where quantum information interacts with complex gauge fields or curved spacetime. By embedding information within the gauge structure, the QFWE enables the development of robust quantum protocols capable of maintaining coherence in dynamic or noisy environments.

The term \(z_{\mu}\)  in the QFWE, which represents the non-local corrections via path-ordered commutators, plays a crucial role in describing quantum correlations and entanglement. Non-locality is central to quantum information, and  \(z_{\mu}\) provides a framework for modeling how entangled states evolve under curved spacetime or strong gravitational fields. This makes the QFWE particularly relevant for studying information flow in extreme environments, such as black holes, where spacetime geometry influences quantum systems.

The QFWE also offers new perspectives on the black hole information paradox, which concerns the apparent loss of information as particles fall into a black hole \cite{42}. Black hole entropy, tied to the entanglement across the event horizon, plays a key role in understanding how quantum information interacts with gravitational systems \cite{43,44}. By modeling the evolution of wavefunctions, entropy, and quantum correlations in these regimes, the QFWE provides a theoretical framework for reconciling the principles of quantum mechanics with general relativity. This approach may contribute to resolving fundamental questions about information preservation and entropy in gravitational systems, offering new insights into one of the most profound problems in modern physics.
\subsection{Quantum Materials}
In condensed matter physics, the QFWE provides a comprehensive framework for understanding quantum materials with exotic properties. It describes emergent phenomena in systems where the collective behavior of particles gives rise to forces and effects that are not present at the microscopic level. For example, the QFWE captures the dynamics of quasi-particles and collective excitations in strongly correlated systems, enabling the study of phenomena such as superconductivity, super-fluidity, and Bose-Einstein condensation.
The QFWE is also crucial for understanding topological quantum materials, such as Weyl semi-metals, which describes manifestations of linear and nonlinear charge-spin conversion \cite{45} and Dirac semi-metals which  are renowned for the host of singular symmetry-protected band degeneracies that can give rise to other exotic phases \cite{46}, where the electronic band structure has a non-trivial topology. These materials exhibit robust surface states, quantized conductance, and unique responses to external fields, making them ideal for applications in quantum computing and spintronics. By modeling the interactions between topology, gauge fields, and quantum dynamics, the QFWE enables the design and discovery of new quantum materials with tailored properties.
\subsection{High-Energy Physics}
The QFWE offers significant potential in high-energy physics by providing a robust framework for studying quantum forces in extreme environments, strong-field regimes, and beyond the Standard Model. Its incorporation of higher-order terms and nonlinear corrections enables the exploration of phenomena at high energies and small spatial scales, where deviations from conventional physics become apparent. The energy-dependent component of the quantum force,  \(E^{2}/hc\), allows the QFWE to model particle interactions in extreme conditions, such as those encountered in high-energy collisions or near black holes and neutron stars, where spacetime curvature plays a critical role.

The QFWE is particularly valuable for modeling particle production in strong electromagnetic fields, such as those created in intense laser-plasma interactions, and for understanding quantum processes in the presence of extreme gravitational fields. In astrophysics, it provides insights into matter under extreme conditions, such as supernovae, gamma-ray bursts, and the interiors of neutron stars, enabling the study of singularities and the quantum structure of spacetime in these environments.
\section{Experimental Feasibility And Observational Signatures}\label{6}
Measuring quantum forces in curved spacetime presents significant challenges owing to the subtle nature of these forces and the extreme conditions required for their observation. Quantum forces emerge from the interplay between a particle’s wavefunction, its momentum, and spacetime curvature, making them difficult to isolate and detect in typical laboratory settings. A primary obstacle is the weakness of quantum-gravitational effects at accessible energy scales, as they are generally expected to become significant only in environments with intense gravitational fields, such as near black holes or during the early universe. 

Despite these challenges, advancements in precision measurement technologies have paved the way for potential laboratory tests. Ultra-cold atoms, trapped ions, and interferometry techniques in microgravity environments, such as satellite-based quantum experiments, are promising tools for probing subtle gravitational effects on quantum particles. For example, cold atom interferometers, which can achieve extreme sensitivity to phase shifts, might detect minute changes in atomic wavefunctions due to weak gravitational fields. This approach has already been explored on the International Space Station (ISS) through experiments such as the Cold Atom Lab, which aims to study quantum gases and Bose-Einstein condensates in microgravity. Extending such setups to specifically test the QFWE can provide experimental insights into quantum forces under low-gravity conditions. 

Matter-wave and optomechanical systems also hold potential for testing predictions from low-energy quantum gravity models \cite{47}, including those described by the QFWE. Optomechanical experiments that use light to manipulate nanomechanical systems can measure extremely small forces with high precision, making them ideal for detecting quantum forces arising from novel gravitational and electromagnetic interactions. For instance, nanomechanical mirrors coupled to optical cavities can detect tiny forces generated by quantum interactions between the mirrors and surrounding fields. Similarly, by observing changes in motion when subjected to controlled gravitational and electromagnetic fields, levitated nanoparticles, such as silica spheres held in optical or magnetic traps, which support an optical whispering-gallery mode and a mechanical mode coupled via an optomechanical interaction \cite{48}, could offer a platform for studying quantum forces. Also, using a linearly polarized laser, we observe the torsional vibration of an optically levitated silica nanodumbbell. This levitated nanodumbbell torsion balance is a novel analog of the Cavendish torsion balance, and provides rare opportunities to observe the Casimir torque and probe the quantum effects of gravity \cite{49}. Furthermore, the phenomenon of nanomechanically induced transparency (NMIT) may be generated from the output probe field in the presence of an effective opto-nanomechanical coupling between the cavity field and the nanosphere, whose steady-state position is influenced by the Coulomb interaction between the cavity mirror and nanosphere \cite{50}. These setups can reveal new force interactions predicted by modified gravity models and electromagnetic gauge theories, such as small but measurable shifts in particle trajectories or forces arising from the coupling between quantum systems, spacetime geometry and electromagnetic fields. By precisely characterizing these forces, researchers can investigate how they can effectively influence one another at the microscopic scale, offering a pathway to test the predictions from the QFWE.

Gravitational wave observatories such as LIGO and Virgo also play an important role by providing data that can shed light on particle behavior near massive astrophysical objects. Although current detectors cannot resolve regions very close to black holes, their sensitivity to gravitational waves is expected to improve significantly over the next few decades \cite{51}. Future detectors, such as the planned Einstein Telescope and space-based Laser Interferometer Space Antenna (LISA), will be sensitive for detecting gravitational waves from sources closer to the event horizons of black holes, possibly revealing subtle quantum corrections in the waveforms from these events. Detecting such corrections would provide indirect evidence of quantum forces in action and offer insights into the QFWE’s predictions. 

Testing gravity also requires predictions from theories beyond general relativity \cite{52}. One way to evaluate modified gravity (MG) theories is to study the large-scale structure (LSS) of the universe \cite{53}. Because cosmic density structures emerge from primordial fluctuations that grow nonlinearly through gravitational collapse, MG theories often produce observable deviations from the predictions of general relativity. This is where the QFWE could contribute by providing a theoretical basis for new predictions about the influence of gravity on quantum systems. As the field of gravitational wave physics advances, the motivation to test alternative gravity theories, including those based on quantum effects, has become increasingly relevant. 

Two emerging observational approaches can probe the innermost regions of black hole spacetimes, offering opportunities to test the QFWE under extreme conditions \cite{54,55}. The first involves detecting gravitational waves from coalescing black holes or extreme-mass-ratio inspirals, where a compact object spirals into a much more massive black hole. Such systems produce gravitational wave signals that are highly sensitive to the strong gravitational fields near the event horizon, potentially revealing quantum corrections predicted by the QFWE. The second approach involves imaging accreting black holes at horizon-scale resolution using the Event Horizon Telescope (EHT), a millimeter-wavelength Very-Long-Baseline Interferometry (VLBI) array. The EHT has already captured the first horizon-scale images of the black hole in the M87 galaxy, and it will soon target the black hole at the center of the Milky Way \cite{56}. These observations could offer indirect tests of the QFWE by revealing how quantum effects might alter the expected behavior of particles near event horizons. 

Relic Gravitational Waves (RGWs), generated in the early universe, provide another promising avenue for probing quantum-gravitational force effects. These waves were studied using Pulsar Timing Arrays (PTAs), which monitor the timing of highly stable millisecond pulsars to detect minute distortions in spacetime and measure low-frequency gravitational waves \cite{57}. For example, PTAs such as the North American Nanohertz Observatory for Gravitational Waves (NANOGrav) \cite{58} and the European Pulsar Timing Array (EPTA) are specifically designed to observe nanohertz-frequency gravitational waves in the range \(f \in \left(10^{-9}, 10^{-7}\right)\) \cite{59}. By examining tiny variations in the arrival times of pulsar signals, researchers can infer the presence of RGWs and compare these signals with predictions from the QFWE. Additionally, deviations in the expected timing residuals—such as phase shifts or subtle modulations, can provide evidence of quantum forces acting on quantum particles in spacetime, as they might alter the propagation of RGWs. Furthermore, cross-correlating pulsar timing data between multiple arrays, such as NANOGrav and the Parkes Pulsar Timing Array (PPTA), can increase the sensitivity to signatures that deviate from classical general relativity. Millisecond pulsars, acting as highly stable cosmic clocks, enable these precise measurements, offering a unique opportunity to detect quantum corrections to spacetime structure and potentially test the predictions of the QFWE.

In laboratory settings, techniques such as atomic force microscopy (AFM) allow scientists to measure the interaction forces at the molecular or ionic level, providing detailed maps of properties such as elasticity, adhesion, and energy dissipation \cite{60,61}. AFM operates using an ultra-sensitive cantilever with a sharp tip that scans the surface of a sample. When the tip interacts with the surface, forces as small as piconewtons cause the cantilever to deflect. These deflections are measured using a laser beam reflected off the cantilever onto a photo detector, allowing the AFM to generate a high-resolution topographic map of the surface. While it is primarily used in materials science, biology, and nanotechnology to study surfaces and interactions at the nanoscale, its exceptional sensitivity has the potential to explore weak quantum forces under specialized experimental conditions. For example, AFM can be employed to investigate the Casimir forces between closely spaced surfaces or nanoparticles, which are quantum forces arising from fluctuations in the electromagnetic field. Also, it can be used to observe the morphology of cells and quantitatively measure their mechanical properties at the nanoscale \cite{62}. Additionally, AFM could be used to measure deviations in expected interaction forces caused by hypothetical modifications to gravity at short distances, as predicted by some quantum gravity models. By scanning surfaces under highly controlled conditions, researchers might detect small quantum-induced forces that differ from classical expectations, providing insight into the behavior of quantum forces in confined geometries. By comparing the predictions of these experimental scenarios to those of the QFWE, its valid would be assessed. 

Additionally, atomic clocks and Bose-Einstein condensates (BECs) under free-fall conditions offer intriguing possibilities for studying tiny gravitational phase shifts. BECs, which are ultra-cold ensembles of atoms in a single quantum state, can exhibit phase coherence making them highly sensitive to gravitational effects. For instance, matter-wave interferometry using BECs can split and recombine atomic wavefunctions to measure phase shifts induced by gravitational fields with extraordinary precision. Such setups have already been used in experiments like the Cold Atom Laboratory (CAL) aboard the International Space Station, where microgravity conditions enhance coherence times and reduce noise. Similarly, atomic clocks, which rely on the precise measurement of atomic energy transitions, can be used in space-based experiments to detect minuscule time dilation effects caused by gravitational fields.

In free-fall setups, such as those implemented in drop towers, parabolic flights, or satellite missions such as MICROSCOPE and the planned STE-QUEST mission, atomic clocks \cite{63} and BEC interferometers could detect subtle shifts in phase or frequency that may reveal quantum forces or deviations from general relativity. For example, measuring differences in the interference fringes of a BEC subjected to a controlled gravitational gradient might uncover small-scale effects from quantum gravity models. Alternatively, dual-species atomic interferometers can compare the gravitational response of different atomic masses to identify deviations attributable to quantum forces or other novel interactions predicted by the QFWE. However, achieving the necessary sensitivity to detect such forces directly remains a formidable challenge, requiring advances in laser cooling, vacuum isolation, and noise reduction technologies. These precision tools offer a pathway to probe the interplay between gravity and quantum mechanics in regimes previously inaccessible. Beyond laboratory experiments, astrophysical and cosmological observations offer more promising avenues for exploring quantum forces in curved spacetime. Near black holes, quantum effects coupled with intense curvature can manifest as Hawking radiation, particle emission, or altered particle trajectories near the event horizon. During cosmic inflation, rapidly evolving spacetime can induce particle creation, potentially leaving imprints in the cosmic microwave background. In extreme environments such as neutron star mergers or supernovae, quantum forces may influence the behavior of matter under high densities and gravitational fields, providing observable signatures that could validate predictions made by the QFWE.

\section{Conclusion}\label{7}
In this study, we introduce the QFWE (Eq. \ref{43}), a novel framework for understanding quantum forces in spacetime. Unlike classical forces, which are largely independent of spacetime geometry, the forces described by the QFWE emerge from the complex interplay between particle wavefunctions, gauge fields, and spacetime curvature. By incorporating covariant derivatives and spin connections, the QFWE provides a unified description of quantum forces that applies across diverse spacetime geometries, from flat to highly curved regions.  This versatility makes the equation suitable for analyzing quantum dynamics in a wide range of physical contexts, from high-energy particle interactions to cosmological and gravitational phenomena. 

Also the modification of the EFE given in Eq. \ref{54} introduces a fundamental new perspective of the nature of spacetime, which is its oscillatory behavior to determine the motion of quantum fields. Unlike conventional approaches in QFT and GR, it suggests a deeper, dynamic relationship where quantum fields actively shape and are shaped by spacetime geometry. One major setback of this framework is that, it does not explicitly quantize the metric thus not making a full quantum gravitational framework. However, it offers a semi-classical approach by treating spacetime as a classical geometric backdrop with an oscillatory nature at the Planck scale. 

This study bridges significant gaps between quantum field theory and general relativity, contributing to the broader quest for a unified theory of fundamental interactions, thus opening new avenues for theoretical exploration and potential experimental or observational testing, particularly in high-energy astrophysical environments and controlled laboratory setups using analog systems.

Future research could focus on experimentally validating the established framework and possibly refine it to replace the metric tensor with its operator, making it a completely quantum gravitational framework. Nevertheless, this paper does not attempt to tackle the problem of a complete "theory of everything" nor replace its candidates.




\bmhead{Attributions}
All the not-referenced results in this work are fully the author’s own contribution.
\bmhead{Funding}
This study was not funded by any organization, institution, or other entity.
\bmhead{Data availability}
Data sharing is not applicable to this article as no datasets were generated or analyzed during the underlying study.

\section*{Declarations}
\bmhead{Conflict of interest}
There are no conflicts of interest to declare that are relevant to the content of this article.
\bmhead{Ethics Statement}
This study adheres to the highest ethical standards for scientific investigations. All theoretical derivations and analyses are original and have been conducted with integrity.





\begin{thebibliography}{50}

\bibitem{1}
J.~Tagg, W.~Reid, and D.~Carlin,
``Schrödinger's Cheshire Cat: A tabletop experiment to measure the Diòsi-Penrose collapse time and demonstrate Objective Reduction (OR),''
\emph{arXiv preprint},
\href{https://arxiv.org/abs/2402.02618}{arXiv:2402.02618}, 2024.

\bibitem{2}
H.~K.~Nguyen,
``Generating variable \(\hbar\) and \(c\) via Fujii-Wetterich model in curved spacetimes,''
\emph{arXiv},
\href{https://arxiv.org/abs/2408.02583}{arXiv:2408.02583}, 2024.

\bibitem{3}
W.~Lucha, D.~Melikhov, and H.~Sazdjian, ``Cluster reducibility of multiquark operators,'' \emph{Phys. Rev. D}, vol.~100, no.~9, p.~094017, \href{https://doi.org/10.1103/PhysRevD.100.094017}{PhysRevD:100.094017}, Nov. 2019.

\bibitem{4}
S.~Mohapatra, S.~Moudgalya, and A.~C.~Balram,
``Exact volume-law entangled eigenstates in a large class of spin models,''
\emph{arXiv preprint}, \href{https://arxiv.org/abs/2410.22773}{arXiv:2410.22773}, 2024.

\bibitem{5}
Z.~Yan, ``Gravitational focusing and horizon entropy for higher-spin fields,'' \emph{arXiv preprint}, \href{https://arxiv.org/abs/2412.07107}{arXiv:2412.07107}, 2024.

\bibitem{6}
C.~Wetterich,
``Cosmology from pregeometry,''
\emph{Phys. Rev. D}, vol.~104, no.~10, p.~104040,
\href{https://doi.org/10.1103/PhysRevD.104.104040}{PhysRevD:104.104040}, 2021.

\bibitem{7}
C.~Wiesendanger,
``Local Lorentz invariance and a new theory of gravitation equivalent to general relativity,''
\emph{Class. Quantum Grav.}, vol.~36, no.~6, p.~065015,
\href{https://doi.org/10.1088/1361-6382/ab04e9}{CQG:36.065015}, 2019.

\bibitem{8}
D.~Flores-Alfonso, ``Non-Abelian black holes in conformal gravity,'' \emph{arXiv preprint}, \href{https://doi.org/10.48550/arXiv.2412.08734}{arXiv:2412.08734}, 2024.

\bibitem{9}
O.~Aharony, O.~Mamroud, S.~Nowik, and M.~Weissman,
``The Bethe Ansatz for the superconformal index with unequal angular momenta,''
\emph{arXiv preprint}, \href{https://arxiv.org/abs/2402.03977}{arXiv:2402.03977}, 2024.

\bibitem{10}
G.~Menezes, ``Leading singularities in higher-derivative Yang–Mills theory and quadratic gravity,'' \emph{Universe}, vol.~8, no.~6, p.~326, \href{https://doi.org/10.3390/universe8060326}{universe8060326}, 2022.

\bibitem{11}
E.~Cinti, C.~Mariani, and M.~Sanchioni,
``The Unbearable Indefiniteness of Spacetime,''
\emph{Found. Phys.}, vol.~55, p.~14,
\href{https://doi.org/10.1007/s10701-025-00819-4}{FoundPhys:55.14}, 2025.

\bibitem{12}
V.~I.~Zhdanov, ``Universal structure of spherically symmetric astrophysical objects in $f(R)$ gravity,'' \emph{arXiv preprint}, \href{https://arxiv.org/abs/2412.03759}{arXiv:2412.03759}, 2024.

\bibitem{13}
C.~Jockel and L.~Menger,
``Effect of torsion on neutron star structure in Einstein-Cartan gravity,''
\emph{Physical Review D}, vol.~110, no.~10, p.~104022,
\href{https://doi.org/10.1103/PhysRevD.110.104022}{PhysRevD:110.104022}, 2024.

\bibitem{14}
A.~R.~P.~Moreira, S.-H.~Dong, and E.~N.~Saridakis, ``Information measures for fermion localization in $f(T, B)$ gravity with non-minimal couplings,'' \emph{arXiv preprint}, \href{https://arxiv.org/abs/2407.15190}{arXiv:2407.15190}, 2024.

\bibitem{15}
A.~A.~Coley, A.~Landry, R.~J.~van den Hoogen, and D.~D.~McNutt,
``Generalized teleparallel de Sitter geometries,''
\emph{The European Physical Journal C}, vol.~83, no.~10, p.~12150,
\href{https://doi.org/10.1140/epjc/s10052-023-12150-1}{EPJC:83.12150}, 2023.

\bibitem{16}
E.~Palti, ``The weak gravity conjecture and scalar fields,'' \emph{J. High Energy Phys.}, vol.~2017, no.~8, p.~034,  \href{http://dx.doi.org/10.1007/JHEP08(2017)034}{JHEP08(2017)034}, 2017.

\bibitem{17}
P.~Berghofer and J.~François, ``Dressing vs. fixing: On how to extract and interpret gauge-invariant content,'' \emph{arXiv preprint}, \href{https://arxiv.org/abs/2404.18582}{arXiv:2404.18582}, 2024.

\bibitem{18}
M.~Donaire, ``Acceleration of a polarized neutron by its weak nuclear self-interaction,'' \emph{arXiv preprint}, \href{https://arxiv.org/abs/2309.02293}{arXiv:2309.02293}, 2023.

\bibitem{19}
R.~E.~Allen,
``The London–Anderson–Englert–Brout–Higgs–Guralnik–Hagen–Kibble–Weinberg mechanism and Higgs boson reveal the unity and future excitement of physics,''
\emph{Journal of Modern Optics}, vol.~61, no.~1, pp.~1–6,
\href{https://doi.org/10.1080/09500340.2013.818170}{09500340.2013.818170}, 2013.

\bibitem{20}
J.~Berger, A.~J.~Long, and J.~Turner, ``Phase of confined electroweak force in the early Universe,'' \emph{Phys. Rev. D}, vol.~100, no.~5, p.~055005, \href{https://doi.org/10.1103/PhysRevD.100.055005}{PhysRevD:100.055005}, 2019.

\bibitem{21}
D.~Ray, ``Some solutions for SU(3) gauge fields,'' \emph{Phys. Rev. D}, vol.~22, no.~8, pp.~2100--2101, \href{https://doi.org/10.1103/PhysRevD.22.2100}{PhysRevD:22.2100}, Oct. 1980.

\bibitem{22}
M.~M.~Ettefaghi and M.~Haghighat, ``Lorentz conserving noncommutative standard model,'' \emph{Phys. Rev. D}, vol.~75, no.~12, p.~125002, \href{https://doi.org/10.1103/PhysRevD.75.125002}{PhysRevD:75.125002}, 2007.

\bibitem{23}
D.~N.~Vollick,
``Liouville Field Theory and Dark Matter,''
\emph{arxiv preprint},
\href{https://arxiv.org/abs/hep-th/0205207}{arXiv:0205207}, 2002.

\bibitem{24}
M.~Cadoni, R.~Casadio, A.~Giusti, and M.~Tuveri, ``Emergence of a dark force in corpuscular gravity,'' \emph{Phys. Rev. D}, vol.~97, no.~4, p.~044047, \href{https://doi.org/10.1103/PhysRevD.97.044047}{PhysRevD:97.044047}, 2018.

\bibitem{25}
S.~Sponar, R.~I.~P.~Sedmik, M.~Pitschmann, H.~Abele, and Y.~Hasegawa,
``Tests of fundamental quantum mechanics and dark interactions with low-energy neutrons,''
\emph{Nature Reviews Physics}, vol.~3, no.~5, pp.~309–327,
\href{https://doi.org/10.1038/s42254-021-00298-2}{s42254-021-00298-2}, 2021.

\bibitem{26}
R.~Lieu,
``The binding of cosmological structures by massless topological defects,''
\emph{Monthly Notices of the Royal Astronomical Society}, vol.~531, no.~1, pp.~1630–1636,
\href{https://doi.org/10.1093/mnras/stae1258}{mnras/stae1258}, 2024.

\bibitem{27}
S.~Akhtar, A.~Manna, and A.~Manu,
``Classical observables using exponentiated spin factors: electromagnetic scattering,''
\emph{J. of High Energy Phys.}, vol.~2024, no.~5, p.~148,
\href{https://doi.org/10.1007/JHEP05(2024)148}{JHEP05(2024)148}, 2024.

\bibitem{28}
J.~Chen, F.~Chen, L.~Yang, Y.~Yang, Z.~Chen, Y.~Wu, Y.~Meng, B.~Yan, X.~Xi, Z.~Zhu, M.~Cheng, G.-G.~Liu, P.~P.~Shum, H.~Chen, R.-G.~Cai, R.-Q.~Yang, Y.~Yang, and Z.~Gao,
``AdS/CFT Correspondence in Hyperbolic Lattices,''
\emph{arXiv}, eprint 2305.04862,
\href{https://arxiv.org/abs/2305.04862}{arxiv:2305.04862}, 2024.

\bibitem{29}
M.~Bojowald, R.~Maartens, and P.~Singh, ``Loop quantum gravity and the cyclic universe,'' \emph{Phys. Rev. D}, vol.~70, no.~8, p.~083517, \href{https://doi.org/10.1103/PhysRevD.70.083517}{PhysRevD:70.083517}, 2004.

\bibitem{30}
C.~Rovelli and S.~Speziale, ``Geometry of loop quantum gravity on a graph,'' \emph{Phys. Rev. D}, vol.~82, no.~4, p.~044018, \href{https://doi.org/10.1103/PhysRevD.82.044018}{PhysRevD:82.044018}, 2010.

\bibitem{31}
W.~Donnelly, ``Entanglement entropy in loop quantum gravity,'' \emph{Phys. Rev. D}, vol.~77, no.~10, p.~104006, \href{https://doi.org/10.1103/PhysRevD.77.104006}{PhysRevD:77.104006}, 2008.

\bibitem{32}
Z.-J.~Liu, J.~Zhou, H.-X.~Meng, X.-Y.~Fan, M.~Xie, F.-L.~Zhang, and J.-L.~Chen, ``Einstein–Podolsky–Rosen steering paradox ‘2=1’ for N qubits,'' \emph{Mod. Phys. Lett. A}, vol.~39, no.~09, p.~2450030, \href{https://doi.org/10.1142/S0217732324500305}{S0217732324500305}, Mar. 2024.

\bibitem{33}
I.~Halder, C.~Vafa, and K.~Xu, ``Black hole entropy for M-theory on the quintic threefold via F-theoretic strings,'' \emph{arXiv preprint}, \href{https://arxiv.org/abs/2404.01380}{arXiv:2404.01380}, 2024.

\bibitem{34}
K.~Yagi and L.~C.~Stein,
``Black hole based tests of general relativity,''
\emph{Classical and Quantum Gravity}, vol.~33, no.~5, p.~054001,
\href{https://doi.org/10.1088/0264-9381/33/5/054001}{0264-9381/33/5/054001}, 2016.

\bibitem{35}
N.~Heidari, A.~A.~Araújo Filho, R.~C.~Pantig, and A.~Övgün, ``Absorption, scattering, geodesics, shadows and lensing phenomena of black holes in effective quantum gravity,'' \emph{arXiv preprint}, \href{https://arxiv.org/abs/2410.08246}{arXiv:2410.08246}, 2024.

\bibitem{36}
V.~P.~Frolov and A.~Zelnikov, \emph{Introduction to Black Hole Physics}, Oxford University Press, \href{https://doi.org/10.1093/acprof:oso/9780199692293.001.0001}{Oxford}, 2018.

\bibitem{37}
N.~Heidari, A.~A.~Araújo Filho, R.~C.~Pantig, and A.~Övgün,
``Absorption, Scattering, Geodesics, Shadows and Lensing Phenomena of Black Holes in Effective Quantum Gravity,''
\emph{arXiv preprint},
\href{https://arxiv.org/abs/2410.08246}{arxiv:2410.08246}, 2024.

\bibitem{38}
S.~W.~Hawking, M.~J.~Perry, and A.~Strominger, ``Soft hair on black holes,'' \emph{Phys. Rev. Lett.}, vol.~116, no.~23, p.~231301, \href{https://doi.org/10.1103/PhysRevLett.116.231301}{PhysRevLett:116.231301}, 2016.

\bibitem{39}
G.~Dvali and C.~Gomez, ``Quantum compositeness of gravity: black holes, AdS and inflation,'' \emph{J. Cosmol. Astropart. Phys.}, vol.~2014, no.~01, p.~023, \href{https://doi.org/10.1088/1475-7516/2014/01/023}{1475-7516/2014/01/023}, 2014.

\bibitem{40}
S.~Carlip, ``Black hole thermodynamics from Euclidean horizon constraints,'' \emph{Phys. Rev. Lett.}, vol.~99, no.~2, p.~021301, \href{https://doi.org/10.1103/PhysRevLett.99.021301}{PhysRevLett:99.021301}, 2007.

\bibitem{41}
D.~Kubizňák and F.~Simovic, ``Thermodynamics of horizons: de Sitter black holes and reentrant phase transitions,'' \emph{Class. Quantum Grav.}, vol.~33, no.~24, p.~245001, \href{https://doi.org/10.1088/0264-9381/33/24/245001}{0264-9381/33/24/245001}, 2016.

\bibitem{42}
S.~Mandal, T.~Parvez, and S.~Shankaranarayanan, ``From Horndeski action to the Callan-Giddings-Harvey-Strominger model and beyond,'' \emph{arXiv preprint}, \href{https://arxiv.org/abs/2311.07921}{arXiv:2311.07921}, 2024.

\bibitem{43}
B.~Chen and Z.-J.~Yin, ``Reflected entropy in an evaporating black hole through non-isometric map,'' \emph{arXiv preprint}, \href{https://arxiv.org/abs/2410.14392}{arXiv:2410.14392}, 2024.

\bibitem{44}
T.~Jacobson, ``Thermodynamics of spacetime: The Einstein equation of state,'' \emph{Phys. Rev. Lett.}, vol.~75, no.~7, pp.~1260–1263, \href{https://doi.org/10.1103/PhysRevLett.75.1260}{PhysRevLett:75.1260}, 1995.

\bibitem{45}
T.~Tang, M.~Li, B.~Lao, X.~Zheng, W.~Zhou, X.~Xu, J.~Pang, Y.-G.~Shi, R.-W.~Li, and Z.~Wang, ``Unveiling the anisotropy of linear and nonlinear charge-spin conversion in Weyl semimetal TaIrTe\textsubscript{4},'' \emph{arXiv preprint}, \href{https://arxiv.org/abs/2411.19062}{arXiv:2411.19062}, 2024.

\bibitem{46}
Y.~Mo, X.~Wang, Z.-Y.~Zhuang, and Z.~Yan, ``Coexistence of chiral Majorana edge states and Bogoliubov Fermi surfaces in two-dimensional nonsymmorphic Dirac semimetal/superconductor heterostructures,'' \emph{arXiv preprint}, \href{https://doi.org/10.48550/arXiv.2411.10851}{arXiv:2411.10851}, 2024.

\bibitem{47}
D.~Carney, P.~C.~E.~Stamp, and J.~M.~Taylor,
``Tabletop experiments for quantum gravity: a user’s manual,''
\emph{Classical and Quantum Gravity}, vol.~36, no.~3, p.~034001,
\href{https://doi.org/10.1088/1361-6382/aaf9ca}{1361-6382/aaf9ca}, 2019.

\bibitem{48}
H.-T.~Li, Z.-Y.~Fan, H.-B.~Zhu, S.~Groeblacher, and J.~Li, ``Microwave-optics entanglement via coupled opto- and magnomechanical microspheres,'' \emph{arXiv preprint}, \href{https://doi.org/10.48550/arXiv.2408.03791}{arXiv:2408.03791}, 2024.

\bibitem{49}
J.~Ahn, Z.~Xu, J.~Bang, Y.-H.~Deng, T.~M.~Hoang, Q.~Han, R.-M.~Ma, and T.~Li,
``Optically Levitated Nanodumbbell Torsion Balance and GHz Nanomechanical Rotor,''
\emph{Phys. Rev. Lett.}, vol.~121, no.~3, p.~033603, 
\href{https://doi.org/10.1103/PhysRevLett.121.033603}{PhysRevLett:121.033603}, 2018.

\bibitem{50}
S.~Shah, A.~Amjad, and H.~Ali, ``Nanomechanically induced transparency in $\mathcal{PT}$-symmetric optical cavities,'' \emph{arXiv preprint}, \href{https://doi.org/10.48550/arXiv.2405.09845}{arXiv:2405.09845}, 2024.

\bibitem{51}
V.~Cardoso and P.~Pani,
``Tests for the existence of black holes through gravitational wave echoes,''
\emph{Nature Astronomy}, vol.~1, no.~9, pp.~586–591,
\href{https://doi.org/10.1038/s41550-017-0225-y}{s41550-017-0225-y}, 2017.

\bibitem{52}
T.~Kobayashi,
``Horndeski theory and beyond: a review,''
\emph{Reports on Progress in Physics}, vol.~82, no.~8, p.~086901,
\href{https://doi.org/10.1088/1361-6633/ab2429}{1361-6633/ab2429}, 2019.

\bibitem{53}
A.~Jiang, W.~Liu, W.~Fang, B.~Li, C.~Barrera-Hinojosa, and Y.~Zhang,
``Minkowski functionals of large-scale structure as a probe of modified gravity,''
\emph{Phys. Rev. D}, vol.~109, no.~8, p.~083537,
\href{https://doi.org/10.1103/PhysRevD.109.083537}{PhysRevD:109.083537}, 2024.

\bibitem{54}
S.~B.~Giddings and D.~Psaltis,
``Event Horizon Telescope observations as probes for quantum structure of astrophysical black holes,''
\emph{Phys. Rev. D}, vol.~97, no.~8, p.~084035,
\href{https://doi.org/10.1103/PhysRevD.97.084035}{PhysRevD:97.084035}, 2018.

\bibitem{55}
E.~Poisson and W.~Israel,
``Internal structure of black holes,''
\emph{Phys. Rev. D}, vol.~41, no.~6, pp.~1796--1809,
\href{https://doi.org/10.1103/PhysRevD.41.1796}{PhysRevD:41.1796}, 1990.

\bibitem{56}
D.~Psaltis,
``Testing general relativity with the Event Horizon Telescope,''
\emph{General Relativity and Gravitation}, vol.~51, no.~10,
\href{http://dx.doi.org/10.1007/s10714-019-2611-5}{s10714-019-2611-5}, 2019.

\bibitem{57}
S.~Z.~Sheikh \emph{et al.}, ``Scintillation bandwidth measurements from 23 pulsars from the AO327 survey,'' \emph{Astrophys. J.}, vol.~976, no.~2, p.~225, \href{https://doi.org/10.3847/1538-4357/ad8659}{1538-4357/ad8659}, 2024.

\bibitem{58}
Q.~Tan, Y.~Wu, and L.~Liu, ``Constraining string cosmology with the gravitational-wave background using the NANOGrav 15-year data set,'' \emph{arXiv preprint}, \href{https://doi.org/10.48550/arXiv.2409.17846}{arXiv:2409.17846}, 2024.

\bibitem{59}
R.~C.~Bernardo and K.-W.~Ng, ``Testing gravity with cosmic variance-limited pulsar timing array correlations,'' \emph{Phys. Rev. D}, vol.~109, no.~10, p.~L101502, \href{https://doi.org/10.1103/PhysRevD.109.L101502}{PhysRevD:109.L101502}, 2024.

\bibitem{60}
R.~Buzio, A.~Gerbi, C.~Bernini, L.~Repetto, and A.~Vanossi,
``Sliding Friction and Superlubricity of Colloidal AFM Probes Coated by Tribo-Induced Graphitic Transfer Layers,''
\emph{Langmuir}, vol.~38, no.~41, pp.~12570–12580,
\href{http://dx.doi.org/10.1021/acs.langmuir.2c02030}{acs.langmuir.2c02030}, 2022.

\bibitem{61}
H.~T.~Çiftçi, M.~Verhage, T.~Cromwijk, L.~Pham Van, B.~Koopmans, K.~Flipse, and O.~Kurnosikov,
``Enhancing sensitivity in atomic force microscopy for planar tip-on-chip probes,''
\emph{Microsystems \& Nanoengineering}, vol.~8, no.~1,
\href{http://dx.doi.org/10.1038/s41378-022-00379-x}{s41378-022-00379-x}, 2022.

\bibitem{62}
A.~S.~Alharbi, M.~S.~Albishi, T.~Maksudov, T.~F.~Alhuwaymel, C.~Aivalioti, K.~S.~AlShebl, N.~R.~Alshamrani, F.~H.~Isikgor, M.~Aldosari, M.~M.~Aljomah, K.~Petridis, T.~D.~Anthopoulos, G.~Kakavelakis, and E.~A.~Alharbi,
``Stable Perovskite Solar Cells via exfoliated graphite as an ion diffusion-blocking layer,''
\emph{arXiv preprint}, \href{https://arxiv.org/abs/2407.21662}{arXiv:2407.21662}, 2024.

\bibitem{63}
P.~Banerjee and D.~Matsakis, ``Introduction to atomic clocks,'' in \emph{An Introduction to Modern Timekeeping and Time Transfer}, Springer Series in Measurement Science and Technology, \href{https://link.springer.com/chapter/10.1007/978-3-031-30780-5_3}{Springer}, Cham, 2023.









\end{thebibliography}
\end{document}